\documentclass[a4paper,fleqn]{cas-sc}

\usepackage[numbers]{natbib}
\usepackage{caption}
\usepackage{subcaption}
\usepackage[verbose]{placeins}
\usepackage{soul}

\usepackage{caption,rotating}
\usepackage{array,siunitx}
\usepackage{float}
\usepackage{booktabs}
\everymath{\displaystyle}

\begin{document}
\let\WriteBookmarks\relax
\def\floatpagepagefraction{1}
\def\textpagefraction{.001}

\shorttitle{Long extensional waves in a layer with irregular free surfaces}
\shortauthors{A. Demirci \& H.T. Uzunoglu}

\title [mode = title]{Nonlinear long extensional waves in an elastic layer with irregular free surfaces}

\author[1]{Ali Demirci}[orcid=0000-0002-9144-5691]
\ead{demircial@itu.edu.tr}
\credit{Conceptualization, Methodology, Formal analysis, Software, Writing -- original draft}

\author[1]{Haldun Taha Uzunoglu}[orcid=0009-0003-4856-7794]
\ead{uzunoglu18@itu.edu.tr}
\cormark[1]
\credit{Methodology, Formal analysis, Validation, Writing -- review \& editing}

\affiliation[1]{organization={Department of Mathematics Engineering, Faculty of Science and Letters, Istanbul Technical University},
     city={Istanbul},
     postcode={34469},
     country={Turkey}}

\cortext[cor1]{Corresponding author}

\begin{abstract}
The propagation of nonlinear long extensional waves in a compressible
hyperelastic layer with irregular free surfaces is studied within the
framework of finite elasticity. The free surfaces are traction-free
and symmetric with respect to the middle plane, and their irregularity
changes the local thickness of the layer by a finite amount over
distances long compared with the wavelength. In a distinguished limit
in which weak nonlinearity, dispersion and the irregularity of the
free surfaces appear at the same order, a multiple-scales analysis of
the two-dimensional boundary-value problem leads to a Korteweg--de
Vries equation with variable coefficients for the longitudinal strain.
The irregularity acts on the wave only through the traction-free
boundary conditions and changes the balance between nonlinearity and
dispersion along the layer. After normalisation, the combined effect
of the material and of the geometry is carried by a single variable
nonlinear coefficient, whose sign decides whether compressive or
tensile solitary waves exist. For special choices of the surface
irregularity, the evolution equation reduces to forms known to be
integrable, namely the Korteweg--de Vries and cylindrical
Korteweg--de Vries equations. For periodically irregular layers of iron
and copper and for model materials, numerical solutions show that a
localized pulse propagates as a solitary wave, splits into a train of
solitary waves or disperses into an oscillatory wave train, depending
on the material, while the irregularity of the free surfaces
redistributes the amplitudes of the emerging solitary waves.
\end{abstract}

\begin{highlights}
\item Long extensional waves in an elastic layer with irregular free surfaces are studied.
\item Surface irregularity alters the balance of nonlinearity and dispersion.
\item Special irregularity profiles reduce the model to known integrable forms.
\item Iron and copper layers support compressive strain solitary waves.
\item Pulses propagate as solitary waves, split into soliton trains or disperse.
\end{highlights}

\maketitle

\section{Introduction}
\label{sec:introduction}
Elastic waveguides such as rods, plates and layers carry long
longitudinal waves over large distances with little attenuation, and
the way these waves change as they travel is a central question in the
mechanics of solids. When the amplitude of the strain is small but
finite, two effects compete. Nonlinearity steepens the wave profile.
It has a geometric part, which comes from the finite-strain kinematics,
and a physical part, which comes from the higher-order elastic
constants. Dispersion spreads the profile. In a waveguide it arises
mainly from the inertia of the transverse motion that the traction-free
lateral surfaces allow. When these two effects balance, the strain can
travel as a solitary wave of permanent form. It was recognised early
that this balance makes the Korteweg--de Vries (KdV) equation and its
generalisations the natural long-wave models of nonlinear elastic
waveguides
\cite{Samsonov1994,Maugin1994,ErofeevKlyueva2002}. Strain solitary
waves in rods and plates have since been studied in detail, including
the influence of a surrounding medium
\cite{Porubov1998}, of two-dimensional effects \cite{Porubov2004} and of
cubic and higher-order terms
\cite{PorubovMaugin2006,PorubovMaugin2008}. Strain solitary waves have
also been generated and recorded in polymer rods \cite{Samsonov1998}, and radiating solitary waves have been
found in bi-layers with delamination
\cite{KhusnutdinovaTranter2017}. These studies show that solitary
waves in solids are not a mathematical idealisation. They are
mechanical objects whose existence and polarity are fixed by the
nonlinear constitutive response of the material, and their potential
use in non-destructive evaluation follows from this.

For a layer with plane, traction-free surfaces, long extensional waves
can be studied by an asymptotic reduction of the two-dimensional
equations of finite elasticity, which leads to a KdV equation whose
nonlinear coefficient depends on the second- and third-order elastic
constants and whose dispersive coefficient reflects the transverse
structure of the fundamental symmetric mode \cite{Teymur2007}.
Boussinesq- and KdV-type models have also been obtained for
compressible hyperelastic plates \cite{Chen2006}. In such models
dispersion is mainly geometric, being associated with the finite
thickness of the waveguide. A different group of studies considers
unbounded media in which dispersion is introduced through the
constitutive law, for instance by higher-gradient, nonlocal or
strain-limiting terms; in this setting, KdV-, Camassa--Holm- and
Kadomtsev--Petviashvili-type equations have been derived for
longitudinal and transverse waves
\cite{DestradeSaccomandi2008,Destrade2011,Erbay2015,Erbay2022,Erbay2024},
and, more recently, the KP equation of plane elastodynamics
\cite{Berjamin2025KP}. The present work belongs to the studies on
plates and layers mentioned first, in which dispersion is of geometric
origin: here the free surfaces of the layer give rise to dispersion, so
that a change in their shape may be expected to affect the balance
between nonlinearity and dispersion.

Real waveguides are rarely uniform. Their thickness changes because of
manufacturing tolerances, wear and corrosion, deliberate tapering or
the natural irregularity of geological layers. The linear theory of
waves in layers with irregular boundaries has a long history in
seismology and elastodynamics. Love waves in layers with irregular
boundaries and in slowly varying layered media were among the first
problems treated
\cite{Wolf1970,Woodhouse1974,MarkenscoffLekoudis1976}. Later work
considered irregular, corrugated and inhomogeneous layers and
half-spaces in increasingly general settings
\cite{Singh2011,Alam2018,Kaur2020,KaurArora2026}, as well as the
scattering of shear waves by irregular surface topographies
\cite{Song2023}. For plate waves, dispersion relations of Rayleigh--Lamb
and shear-horizontal waves have been obtained for corrugated
waveguides \cite{TavafBanerjee2020}. The propagation of Lamb waves in a
layer with irregular free surfaces and the effect of surface roughness on
generalised Rayleigh waves have also been examined
\cite{Ahmetolan2023,Sezer2026}. The common conclusion of these studies
is that the geometry of the boundaries changes the dispersive
properties of the waveguide. Much less is known when nonlinearity is
also present. Nonlinear surface SH waves in a half-space covered by an
irregular layer obey an evolution equation with variable coefficients
\cite{DeliktasTeymur2022}. The focusing of strain solitary waves in a
narrowing rod has been demonstrated experimentally and explained by
the variation of the coefficients of the governing equation along the
rod \cite{Samsonov1998}. The nonlinear S0 Lamb mode in a plate of
linearly varying thickness \cite{Hu2019} and solitary waves in rods
made of an inhomogeneous hyperelastic material \cite{DaiHuo2002} show
the same sensitivity to longitudinal variations of the waveguide.

When the properties of a waveguide vary slowly compared with the
wavelength, the natural long-wave model is a KdV equation with variable
coefficients. Its theory has been developed mainly for water waves and
internal waves over variable topography. A solitary wave in a slowly
varying medium adjusts adiabatically: its amplitude follows from the
conservation of the wave action, or energy flux, and a trailing shelf
of small amplitude is generated behind it. If the medium changes
rapidly, or too much, the solitary wave fissions into several solitary
waves together with a dispersive tail
\cite{ElGrimshaw2002,GrimshawPudjaprasetya2004}. Whether elevation or
depression waves emerge is decided by the sign of the nonlinear
coefficient, which may change along the path of propagation
\cite{GrimshawYuan2016,Zhi2021}. The variable-coefficient KdV equation
reduces to forms known to be integrable only for particular choices of
its coefficients; the Painlev\'e analysis yields conditions on the
coefficients that are satisfied, for instance, by the
constant-coefficient and cylindrical KdV equations
\cite{Tian2006,Zhang2006,TianZhang2014}. A layer of variable thickness
is the elastic counterpart of fluid flow over variable topography. In
the solid, however, the coefficients of the evolution equation are not
free parameters. They are fixed by the finite-elastic constitutive law
and by the traction-free condition on the irregular free surfaces, and how the
geometry enters each coefficient has to be found from the
boundary-value problem itself.

The purpose of the present paper is to answer this question for long
extensional waves in a compressible hyperelastic layer whose
traction-free surfaces are irregular and symmetric with respect to the
middle plane. The thickness is allowed to change by a finite amount on
a scale long compared with the wavelength. We adopt a distinguished
limit in which weak nonlinearity, dispersion and the slow variation of
the thickness appear at the same order. The method of multiple scales
applied to the full two-dimensional problem of finite elasticity then
gives a KdV equation with variable coefficients for the longitudinal
strain. The analysis shows that the irregularity enters this equation
in only two ways. The first is a linear term that expresses the
conservation of the energy flux through a cross-section. The second is
a dispersive coefficient proportional to the square of the local
thickness. The nonlinear coefficient is that of the uniaxial response
of a layer with free surfaces, and the dispersive coefficient depends on
Poisson's ratio alone. After normalisation, a single variable
coefficient governs the polarity of the solitary waves. For special
choices of the surface irregularity, namely a uniform layer and a
cube-root thickness profile, the equation reduces to forms known to be
integrable, the KdV and the cylindrical KdV equations. Bounded
irregularities do not lead to such forms, and the evolution is then
computed numerically. The evolution of localized pulses in periodically irregular
layers of iron and copper, and of model materials, is computed and
interpreted in terms of the material and of the geometry.

The paper is organised as follows. The boundary-value problem is
formulated in Section~\ref{sec:formulation}. The asymptotic analysis
leading to the variable-coefficient KdV equation, and the mechanical
interpretation of its coefficients, are given in
Section~\ref{sec:multiple_scales}. Section~\ref{sec:soliton_evolution}
deals with the normalised equation, the polarity of the solitary
waves, integrability and the numerical results. Conclusions and
possible extensions are given in Section~\ref{sec:conclusions}. The
stress components and the operators of the perturbation problems are
listed in the Appendices.

\section{Formulation of the problem}
\label{sec:formulation}

In this section the boundary-value problem governing the plane motion
of an elastic layer with irregular free surfaces is set up within the theory of
finite elasticity. The geometry of the layer, its motion and the
balance equations are introduced first, in the material description.
The constitutive equations appropriate to deformations of small but
finite amplitude are given next. Finally, the problem is written in
dimensionless form, and the small parameters that characterise long
waves, weak nonlinearity and the slow variation of the thickness are
identified.

\subsection{Geometry, motion and balance equations}
\label{subsec:geometry}

Consider a homogeneous, isotropic and compressible hyperelastic layer
referred to a fixed rectangular Cartesian frame. The material
coordinates of a particle are denoted by $X_K=(X,Y,Z)$ and its spatial
coordinates at time $t$ by $x_k$. In the undeformed configuration the
layer occupies the region
\begin{equation}
\mathcal{B}_{0}
=
\left\{
(X,Y,Z):\;
-H(X)<Y<H(X),\quad
-\infty<X,Z<\infty
\right\},
\label{eq:reference_configuration}
\end{equation}
bounded by the two material surfaces
\begin{equation}
\mathcal{S}^{\pm}_{0}:\qquad Y=\pm H(X),
\qquad
H(X)=h+a f_1 .
\label{eq:dimensional_geometry}
\end{equation}
Here $2h$ is the thickness of the corresponding layer with plane free surfaces, $a$ is the amplitude of the surface irregularity and $f_1$ is a
bounded, sufficiently smooth function of $X$ whose longitudinal scale
is specified in Section~\ref{subsec:dimensionless}. The two free surfaces are
mirror images of each other with respect to the middle plane $Y=0$.
This restriction is essential for what follows: it allows the
symmetric (extensional) mode of the uniform layer to be continued into
a waveguide whose local thickness $2H(X)$ varies along the direction of
propagation.

The motion is restricted to the $(X,Y)$-plane,
\begin{equation}
x_1=X+u_1(X,Y,t),
\qquad
x_2=Y+u_2(X,Y,t),
\qquad
x_3=Z,
\label{eq:motion}
\end{equation}
where $u_k$ are the components of the displacement vector. In the
absence of body forces, the balance of linear momentum referred to the
undeformed configuration reads \cite{EringenSuhubi1975}
\begin{equation}
T_{Kk,K}=\rho_0\,\ddot{u}_k,
\qquad k=1,2,
\label{eq:balance_momentum}
\end{equation}
where $T_{Kk}$ is the first Piola--Kirchhoff stress tensor, $\rho_0$ is
the mass density in the undeformed configuration, a comma followed by
a majuscule index denotes partial differentiation with respect to the
corresponding material coordinate, a superposed dot denotes the
material time derivative, and summation over repeated indices is
understood.

The surfaces $\mathcal{S}^{\pm}_{0}$ are free of traction. Denoting by
$N^{\pm}_K$ the components of the exterior unit normals of
$\mathcal{S}^{\pm}_{0}$,
\begin{equation}
\left(N^{\pm}_1,N^{\pm}_2\right)
=
\frac{\left(\mp H'(X),\,\pm1\right)}{\sqrt{1+H'(X)^2}},
\label{eq:surface_normals}
\end{equation}
the boundary conditions are \cite{EringenSuhubi1975}
\begin{equation}
N^{\pm}_{K}T_{Kk}=0,
\qquad k=1,2,
\qquad\text{on }\ \mathcal{S}^{\pm}_{0}.
\label{eq:traction_free}
\end{equation}
In contrast to a layer of uniform thickness, \eqref{eq:traction_free}
involves a longitudinally varying normal. The irregular geometry
therefore enters the boundary-value problem through the traction
vector itself, without any external surface loading, and this is the
only mechanism by which the variable thickness acts on the wave field.

\subsection{Constitutive equations}
\label{subsec:constitutive}

The balance equations \eqref{eq:balance_momentum} and the boundary
conditions \eqref{eq:traction_free} hold for any material; the mechanical response of the layer is
specified by the constitutive equations. For a hyperelastic material,
the stress is derived from a strain energy function $\Sigma$, measured per unit volume of the undeformed
configuration, so that \cite{EringenSuhubi1975}
\begin{equation}
T_{Kk}=\frac{\partial\Sigma}{\partial x_{k,K}} .
\label{eq:hyperelastic}
\end{equation}
For an isotropic solid, $\Sigma$ depends on the deformation through the
principal invariants $I_E$, $II_E$ and $III_E$ of the Lagrangian strain
tensor
\begin{equation}
E_{KL}
=
\frac{1}{2}
\left(
u_{K,L}+u_{L,K}+u_{M,K}u_{M,L}
\right),
\qquad
u_K=u_k\delta_{kK}.
\label{eq:lagrangian_strain}
\end{equation}
We consider deformations of small but finite amplitude and accordingly
expand $\Sigma$ in powers of the strain invariants $I_E$, $II_E$ and
$III_E$, retaining all terms up to the third degree in the strain
components \cite{EringenSuhubi1975,Teymur2007}. The resulting
expression involves the Lam\'e constants $\lambda$ and $\mu$ and the
third-order elastic constants $l$, $m$ and $n$. Since $III_E$ vanishes
identically for the plane motion \eqref{eq:motion}, the constant $n$
does not enter the problem.

Substitution of \eqref{eq:lagrangian_strain} and of the strain energy
function described above into \eqref{eq:hyperelastic} gives the stress
components as polynomials in the displacement gradients. Consistently with the degree of the strain energy
function, they are truncated at the second degree;
the resulting expressions are listed in
Appendix~\ref{app:constitutive}. They involve the two linear wave
speeds
\begin{equation}
c_L^2=\frac{\lambda+2\mu}{\rho_0},
\qquad
c_T^2=\frac{\mu}{\rho_0},
\label{eq:wave_speeds}
\end{equation}
and four nonlinear material coefficients
\begin{equation}
\eta_i=\frac{\beta_i}{\rho_0},
\qquad i=1,\ldots,4,
\label{eq:eta_definition}
\end{equation}
with
\begin{equation}
\beta_1=\frac{3}{2}\lambda+3\mu+3l,
\qquad
\beta_2=\frac{1}{2}\lambda+3l+m,
\qquad
\beta_3=\mu-\frac{m}{2},
\qquad
\beta_4=\frac{\lambda}{2}+\mu-\frac{m}{4}.
\label{eq:beta_definition}
\end{equation}
Each $\beta_i$ combines a contribution of the geometric nonlinearity,
carried by the Lam\'e constants, with a contribution of the physical
nonlinearity, carried by the third-order constants. The coefficients
$\eta_i$ have the dimension of a squared velocity. It will be seen that
the linear constants and the nonlinear coefficients enter the
long-wave model in clearly separated ways.

\subsection{Dimensionless boundary-value problem}
\label{subsec:dimensionless}

Equations \eqref{eq:balance_momentum} and \eqref{eq:traction_free},
together with the stress components of
Appendix~\ref{app:constitutive}, constitute a nonlinear
boundary-value problem for $u_1$ and $u_2$ in a domain of variable
width. Its asymptotic treatment requires a scaling that distinguishes
the direction of propagation from the transverse direction.
Let $L$ be a characteristic wavelength in the direction of propagation
and $c$ a characteristic velocity, to be fixed in the course of the
analysis. We introduce the dimensionless quantities
\begin{equation}
x=\frac{X}{L},
\qquad
y=\frac{Y}{h},
\qquad
\tau=\frac{c\,t}{L},
\qquad
u=\frac{u_1}{L},
\qquad
v=\frac{u_2}{h},
\qquad
A=\frac{a}{h},
\qquad
\delta=\frac{h}{L}.
\label{eq:dimensionless_variables}
\end{equation}
The longitudinal and transverse coordinates are scaled differently, and
this anisotropy must be kept when the displacement gradients are
transformed:
\begin{equation}
u_{1,X}=\frac{\partial u}{\partial x},
\qquad
u_{1,Y}=\frac{1}{\delta}\frac{\partial u}{\partial y},
\qquad
u_{2,X}=\delta\,\frac{\partial v}{\partial x},
\qquad
u_{2,Y}=\frac{\partial v}{\partial y}.
\label{eq:scaled_gradients}
\end{equation}
Since the dimensional displacement components $u_1$ and $u_2$ do not
appear again, subscripts attached to $u$ and $v$ will from now on
indicate the order of a term in a perturbation expansion.

We are concerned with waves that are long compared with the thickness
of the layer, $\delta\ll1$, and of small but finite amplitude. Let
$\varepsilon\ll1$ measure the magnitude of the displacement gradients.
The effects of nonlinearity and of dispersion are of the same order in
the distinguished limit $\delta^2=O(\varepsilon)$
\cite{Teymur2007}, and without loss of
generality we set
\begin{equation}
\delta^2=\varepsilon .
\label{eq:epsilon_delta_relation}
\end{equation}
The surface irregularity is assumed to vary slowly on the scale of the
wavelength; specifically, $f_1$ depends on $X$ only through the slow
variable
\begin{equation}
x_1=\varepsilon x .
\label{eq:slow_variable}
\end{equation}
In dimensionless form, the free surfaces of the layer are $y=\pm P(x_1)$, with
\begin{equation}
P(x_1)=\frac{H}{h}=1+A f_1(x_1)>0,
\qquad
P'(x_1)=\frac{dP}{dx_1}=A\frac{df_1}{dx_1},
\label{eq:P_definition}
\end{equation}
so that
\begin{equation}
\frac{\partial P}{\partial x}=\varepsilon P'(x_1),
\qquad
P=O(1),
\qquad
P'=O(1).
\label{eq:boundary_derivative_scaling}
\end{equation}
No smallness is imposed on the amplitude $A$ of the irregularity: the
asymptotic assumption concerns the rate at which the local thickness
changes along the layer, not the magnitude of the change. The geometry
can therefore modify the waveguide by a finite amount, while its effect
on the wave accumulates gradually over the long propagation distance
$x=O(\varepsilon^{-1})$.

With \eqref{eq:dimensionless_variables}--\eqref{eq:epsilon_delta_relation}
and the stress components of Appendix~\ref{app:constitutive}, the
balance equations \eqref{eq:balance_momentum} take the form
\begin{equation}
\mathcal{L}_1(u,v)=\mathcal{N}_1(u,v),
\qquad
\mathcal{L}_2(u,v)=\mathcal{N}_2(u,v),
\qquad
-P(x_1)<y<P(x_1),
\label{eq:dimensionless_field_problem}
\end{equation}
where $\mathcal{L}_i$ are linear and $\mathcal{N}_i$ are quadratic
differential operators. The normals \eqref{eq:surface_normals} become,
up to a positive factor,
\begin{equation}
\left(N^{\pm}_1,N^{\pm}_2\right)
\ \parallel\
\left(\mp\delta\,\frac{\partial P}{\partial x},\ \pm1\right),
\label{eq:dimensionless_normals}
\end{equation}
so that the inclination of the free surfaces is weighted by the slenderness
parameter, and the traction conditions \eqref{eq:traction_free} are
written as
\begin{equation}
\mathcal{B}^{\pm}_1(u,v)=0,
\qquad
\mathcal{B}^{\pm}_2(u,v)=0
\qquad\text{on }\ y=\pm P(x_1),
\label{eq:dimensionless_bc}
\end{equation}
where the subscripts $1$ and $2$ refer to the $x$- and $y$-components
of the traction. The explicit forms
of $\mathcal{L}_i$, $\mathcal{N}_i$ and $\mathcal{B}^{\pm}_i$ are given
in Appendix~\ref{app:operators}.

Because the geometry is symmetric with respect to $y=0$, the
boundary-value problem
\eqref{eq:dimensionless_field_problem}--\eqref{eq:dimensionless_bc}
admits solutions with
\begin{equation}
u(x,-y,\tau)=u(x,y,\tau),
\qquad
v(x,-y,\tau)=-v(x,y,\tau),
\label{eq:symmetry_conditions}
\end{equation}
which constitute the symmetric (extensional) family considered in this
work. For such solutions the conditions on $y=-P$ are satisfied
whenever those on $y=P$ are, and only the latter need be imposed. In the
limit $A\rightarrow0$ the problem reduces to that of the symmetric
long waves in a layer of uniform thickness.

\section{Asymptotic analysis}
\label{sec:multiple_scales}

The dimensionless boundary-value problem of
Section~\ref{subsec:dimensionless} is now solved asymptotically for relatively small values of
$\varepsilon$, $0<\varepsilon\ll1$. The expansion is introduced first. The
perturbation problems are then solved successively: the first two
orders determine the transverse structure of the wave and its velocity
of propagation, whereas the third order, at which nonlinearity,
dispersion and the variation of the thickness appear simultaneously,
provides the equation governing the evolution of the wave amplitude.
The coefficients of this equation are discussed at the end of the
section.

\subsection{Multiple-scales expansion}
\label{subsec:expansion}

The propagation of long waves of small but finite amplitude is studied
by the method of multiple scales \cite{Jeffrey1994}. In addition to the slow variable
\eqref{eq:slow_variable} we write $x_0=x$ and regard the field
quantities as functions of $(x_0,x_1,y,\tau)$, so that
\begin{equation}
\frac{\partial}{\partial x}
\longrightarrow
\frac{\partial}{\partial x_0}
+\varepsilon\frac{\partial}{\partial x_1}.
\label{eq:ms_derivative_expansion}
\end{equation}
The displacement components are expanded as
\begin{equation}
u=\sum_{n=1}^{\infty}\varepsilon^n u_n(x_0,x_1,y,\tau),
\qquad
v=\sum_{n=1}^{\infty}\varepsilon^n v_n(x_0,x_1,y,\tau).
\label{eq:ms_expansion}
\end{equation}
Substituting \eqref{eq:ms_derivative_expansion} and
\eqref{eq:ms_expansion} into
\eqref{eq:dimensionless_field_problem}--\eqref{eq:dimensionless_bc}
and collecting the coefficients of like powers of $\varepsilon$ yields
a hierarchy of boundary-value problems on the strip $-P(x_1)<y<P(x_1)$,
in which $x_1$ enters only as a parameter. With the operators written as
in Appendix~\ref{app:operators}, the problem of order $\varepsilon^n$
determines the dependence of $u_n$ and $v_{n-1}$ on $y$. The operators
$\mathcal{L}^{nj}$, $\mathcal{N}^{nj}$ and $\mathcal{B}^{nj}$
$(j=1,2)$ appearing below are listed in
Appendix~\ref{app:order_operators}. In view of
\eqref{eq:symmetry_conditions}, $u_n$ is even and $v_n$ is odd in $y$,
and the boundary conditions are imposed on $y=P(x_1)$ only.

\subsection{First- and second-order problems}
\label{subsec:first_second}

We begin with the two lowest orders of the hierarchy. Their solutions
describe a wave that is uniform across the thickness to leading order,
and their compatibility fixes the characteristic velocity $c$.

\medskip
\noindent\textbf{$O(\varepsilon)$:}
The first-order problem is
\begin{equation}
c_T^2\frac{\partial^2u_1}{\partial y^2}=0,
\qquad
c_T^2\frac{\partial u_1}{\partial y}=0
\quad\text{on } y=P(x_1),
\label{eq:ms_Oeps_1}
\end{equation}
whose even solution is
\begin{equation}
u_1=A^{*}(x_0,x_1,\tau),
\label{eq:ms_first_order_u_final}
\end{equation}
where $A^{*}$ is an as yet arbitrary function.

\medskip
\noindent\textbf{$O(\varepsilon^2)$:}
The second-order problem is
\begin{align}
-c_T^2\frac{\partial^2u_2}{\partial y^2}
+\mathcal{L}^{21}(u_1,v_1)
&=\mathcal{N}^{21}(u_1,v_1),
\label{eq:ms_Oeps2_u}
\\
\mathcal{L}^{22}(u_1,v_1)
&=\mathcal{N}^{22}(u_1,v_1),
\label{eq:ms_Oeps2_v}
\end{align}
with
\begin{align}
c_T^2\frac{\partial u_2}{\partial y}
+\mathcal{B}^{21}(u_1,v_1)&=0,
\label{eq:ms_Oeps2_bc_1}
\\
\mathcal{B}^{22}(u_1,v_1)&=0
\label{eq:ms_Oeps2_bc_2}
\end{align}
on $y=P(x_1)$. By \eqref{eq:ms_first_order_u_final}, equation
\eqref{eq:ms_Oeps2_v} reduces to
$c_L^2\,\partial^2v_1/\partial y^2=0$, whose odd solution is
$v_1=\gamma_1(x_0,x_1,\tau)\,y$. The condition
\eqref{eq:ms_Oeps2_bc_2} then gives
$c_L^2\gamma_1+\left(c_L^2-2c_T^2\right)\partial A^{*}/\partial x_0=0$,
that is,
\begin{equation}
v_1
=
\left(\frac{2c_T^2}{c_L^2}-1\right)
\frac{\partial A^{*}}{\partial x_0}\,y .
\label{eq:ms_first_order_v_final}
\end{equation}
Thus, to leading order, each cross-section contracts or expands
uniformly in the transverse direction in proportion to the local
longitudinal strain, in such a way that the normal traction on the
free surfaces vanishes.

Inserting \eqref{eq:ms_first_order_u_final} and
\eqref{eq:ms_first_order_v_final} into \eqref{eq:ms_Oeps2_u} we obtain
\begin{equation}
\frac{\partial^2u_2}{\partial y^2}=V^{*}(x_0,x_1,\tau),
\qquad
V^{*}
=
\frac{c^2}{c_T^2}\frac{\partial^2A^{*}}{\partial\tau^2}
+\left(\frac{2c_T^2}{c_L^2}-3\right)
\frac{\partial^2A^{*}}{\partial x_0^2},
\label{eq:ms_Vstar_definition}
\end{equation}
and hence the even solution
\begin{equation}
u_2=V^{*}(x_0,x_1,\tau)\,\frac{y^2}{2}+B_1(x_0,x_1,\tau),
\label{eq:ms_second_order_u_general}
\end{equation}
where $B_1$ is a new arbitrary function. The remaining boundary
condition \eqref{eq:ms_Oeps2_bc_1} becomes
\begin{equation}
P(x_1)\,c_T^2
\left\{
\frac{c^2}{c_T^2}\frac{\partial^2A^{*}}{\partial\tau^2}
-4\left(1-\frac{c_T^2}{c_L^2}\right)
\frac{\partial^2A^{*}}{\partial x_0^2}
\right\}=0 .
\label{eq:ms_second_order_boundary_reduced}
\end{equation}
Since $P>0$, the expression in braces must vanish. This is the
solvability condition of the second-order problem. We now fix the
characteristic velocity as
\begin{equation}
c^2
=
4c_T^2\left(1-\frac{c_T^2}{c_L^2}\right)
=
\frac{E}{\left(1-\nu^2\right)\rho_0},
\label{eq:ms_characteristic_velocity}
\end{equation}
where $E$ is Young's modulus and $\nu$ is Poisson's ratio; $c$ is the
velocity of long extensional waves in a plate \cite{EringenSuhubi1975}. The solvability
condition then reduces to
\begin{equation}
\frac{\partial^2A^{*}}{\partial\tau^2}
-\frac{\partial^2A^{*}}{\partial x_0^2}=0 ,
\label{eq:ms_second_order_wave_equation_reduced}
\end{equation}
and for a wave travelling in the positive $x_0$-direction we take
\begin{equation}
A^{*}=A^{*}(\eta,\xi),
\qquad
\eta=\tau-x_0,
\qquad
\xi=x_1 .
\label{eq:travelling_coordinate}
\end{equation}
It is worth noting that the local thickness $P$ factors out of
\eqref{eq:ms_second_order_boundary_reduced}: the velocity of the long
wave is not affected by the surface irregularity, whose influence is
deferred to the next order. The general forms
\eqref{eq:ms_first_order_u_final}--\eqref{eq:ms_second_order_u_general},
in which the derivatives with respect to $\tau$ and $x_0$ are kept
separate, are used in the following subsection; the relations
\eqref{eq:ms_characteristic_velocity} and
\eqref{eq:travelling_coordinate} are imposed only in the final
solvability condition.

\subsection{Third-order problem}
\label{subsec:third_order}

The amplitude $A^{*}$ is left undetermined by the first two orders
apart from its dependence on $\tau-x_0$. Its dependence on the slow
variable $x_1$ follows from the next order, to which we now turn.

\medskip
\noindent\textbf{$O(\varepsilon^3)$:}
The third-order problem is
\begin{align}
-c_T^2\frac{\partial^2u_3}{\partial y^2}
+\mathcal{L}^{31}(u_1,u_2,v_1,v_2)
&=\mathcal{N}^{31}(u_1,u_2,v_1,v_2),
\label{eq:ms_Oeps3_u}
\\
\mathcal{L}^{32}(u_1,u_2,v_1,v_2)
&=\mathcal{N}^{32}(u_1,u_2,v_1,v_2),
\label{eq:ms_Oeps3_v}
\end{align}
with
\begin{align}
c_T^2\frac{\partial u_3}{\partial y}
+\mathcal{B}^{31}(u_1,u_2,v_1,v_2)&=0,
\label{eq:ms_Oeps3_bc_1}
\\
\mathcal{B}^{32}(u_1,u_2,v_1,v_2)&=0
\label{eq:ms_Oeps3_bc_2}
\end{align}
on $y=P(x_1)$. Using the lower-order solutions in
\eqref{eq:ms_Oeps3_v} we find
\begin{equation}
\frac{\partial^2v_2}{\partial y^2}=\chi(x_0,x_1,\tau)\,y,
\qquad
\chi
=
\frac{c^2}{c_L^2}
\left(\frac{2c_T^2}{c_L^2}-\frac{c_L^2}{c_T^2}\right)
\frac{\partial^3A^{*}}{\partial x_0\partial\tau^2}
+\left(3-\frac{4c_T^2}{c_L^2}\right)
\frac{\partial^3A^{*}}{\partial x_0^3},
\label{eq:chi_definition}
\end{equation}
and therefore the odd solution
\begin{equation}
v_2=\chi(x_0,x_1,\tau)\,\frac{y^3}{6}+K_1(x_0,x_1,\tau)\,y .
\label{eq:ms_Oeps3_v2_solution}
\end{equation}
The function $K_1$ is determined by the boundary condition
\eqref{eq:ms_Oeps3_bc_2}:
\begin{align}
K_1
={}&
\frac{P^2}{2}
\left(1-\frac{c_T^2}{c_L^2}\right)
\left\{
\frac{2c^2}{c_L^2}
\frac{\partial^3A^{*}}{\partial x_0\partial\tau^2}
-\frac{4c_T^2}{c_L^2}
\frac{\partial^3A^{*}}{\partial x_0^3}
\right\}
\nonumber\\
&-
\frac{1}{c_L^2}
\left\{
\eta_1\left(\frac{2c_T^2}{c_L^2}-1\right)^2
+2\eta_2\left(\frac{2c_T^2}{c_L^2}-1\right)
+\eta_2
\right\}
\left(\frac{\partial A^{*}}{\partial x_0}\right)^2
+\left(\frac{2c_T^2}{c_L^2}-1\right)
\left\{
\frac{\partial A^{*}}{\partial x_1}
+\frac{\partial B_1}{\partial x_0}
\right\}.
\label{eq:ms_Oeps3_K1}
\end{align}
The three groups of terms in \eqref{eq:ms_Oeps3_K1} represent,
respectively, the correction of the transverse contraction due to the
non-uniform distribution of the displacement across the thickness,
the second-order (nonlinear) part of the transverse contraction, and
the contraction associated with the slow modulation and with the
second-order longitudinal displacement.

Next, the solutions for $u_1$, $v_1$, $u_2$ and $v_2$ are substituted
into \eqref{eq:ms_Oeps3_u}. This gives
\begin{equation}
\begin{aligned}
\frac{\partial^2u_3}{\partial y^2}
={}&
\frac{y^2}{2}
\Bigg[
\frac{c^4}{c_T^4}
\frac{\partial^4A^{*}}{\partial\tau^4}
-\frac{2c^2\left(2c_L^4-c_T^4\right)}{c_L^4c_T^2}
\frac{\partial^4A^{*}}{\partial x_0^2\partial\tau^2}
+\left(5-\frac{4c_T^2}{c_L^2}\right)
\frac{\partial^4A^{*}}{\partial x_0^4}
\Bigg]
\\
&+
\left(\frac{4c_T^2}{c_L^2}-6\right)
\frac{\partial^2A^{*}}{\partial x_0\partial x_1}
+P^2\,
\frac{\left(c_L^2-c_T^2\right)^2}{c_L^4c_T^2}
\left(
-c^2\frac{\partial^4A^{*}}{\partial x_0^2\partial\tau^2}
+2c_T^2\frac{\partial^4A^{*}}{\partial x_0^4}
\right)
\\
&-
\frac{4c^2\eta_4}{c_L^2c_T^2}
\frac{\partial A^{*}}{\partial x_0}
\frac{\partial^2A^{*}}{\partial\tau^2}
+Z^{*}
\frac{\partial A^{*}}{\partial x_0}
\frac{\partial^2A^{*}}{\partial x_0^2}
+\frac{c^2}{c_T^2}\frac{\partial^2B_1}{\partial\tau^2}
+\left(\frac{2c_T^2}{c_L^2}-3\right)
\frac{\partial^2B_1}{\partial x_0^2},
\end{aligned}
\label{eq:ms_Oeps3_u3_equation}
\end{equation}
where the coefficient of the remaining quadratic terms is
\begin{equation}
Z^{*}
=
\frac{2\eta_1\left(8c_L^2c_T^2-5c_L^4-4c_T^4\right)}{c_L^6}
+\frac{2\eta_2\left(5c_L^2-8c_T^2\right)}{c_L^4}
+\frac{2\eta_3\left(c_L^2-2c_T^2\right)}{c_L^4}
+\frac{4\eta_4\left(3c_L^2-2c_T^2\right)}{c_L^4}.
\label{eq:Z_star}
\end{equation}
An explicit expression for $u_3$ is not needed. Since $u_3$ is even in
$y$, $\partial u_3/\partial y$ vanishes on $y=0$; integrating
\eqref{eq:ms_Oeps3_u3_equation} with respect to $y$ from $0$ to
$P(x_1)$ and eliminating $\partial u_3/\partial y$ on $y=P$ by means of
the boundary condition \eqref{eq:ms_Oeps3_bc_1}, we obtain
the solvability condition of the third-order problem,
\begin{equation}
\begin{aligned}
&P\,\Bigg[
\frac{c^2}{c_T^2}\frac{\partial^2B_1}{\partial\tau^2}
+\frac{4\left(c_T^2-c_L^2\right)}{c_L^2}
\frac{\partial^2B_1}{\partial x_0^2}
+8\left(\frac{c_T^2}{c_L^2}-1\right)
\frac{\partial^2A^{*}}{\partial x_0\partial x_1}
\\
&\qquad
+\frac{-16c_T^4\eta_1
+24c_L^2c_T^2\left(\eta_1-\eta_2\right)
+12c_L^4\left(\eta_2-\eta_1\right)}{c_L^6}\,
\frac{\partial A^{*}}{\partial x_0}
\frac{\partial^2A^{*}}{\partial x_0^2}
\Bigg]
\\
&+P^3\,\Bigg\{
\frac{c^4}{6c_T^4}
\frac{\partial^4A^{*}}{\partial\tau^4}
-\frac{c^2\left(11c_L^4-18c_L^2c_T^2+8c_T^4\right)}{6c_L^4c_T^2}
\frac{\partial^4A^{*}}{\partial x_0^2\partial\tau^2}
+\frac{2\left(c_L^2-c_T^2\right)\left(5c_L^2-6c_T^2\right)}{3c_L^4}
\frac{\partial^4A^{*}}{\partial x_0^4}
\Bigg\}
\\
&+4\left(\frac{c_T^2}{c_L^2}-1\right)
P'\,\frac{\partial A^{*}}{\partial x_0}=0 .
\end{aligned}
\label{eq:third_order_solvability_condition}
\end{equation}
The coefficients $\eta_3$ and $\eta_4$ do not appear in
\eqref{eq:third_order_solvability_condition}: the terms proportional to
them in \eqref{eq:ms_Oeps3_u3_equation} are balanced exactly by the
corresponding terms of $\mathcal{B}^{31}$. The last term of
\eqref{eq:third_order_solvability_condition} is the only one in which
the inclination of the free surfaces enters; it originates from the term of
$\mathcal{B}^{31}$ that is proportional to $P'$.

\subsection{Variable-coefficient Korteweg--de Vries equation}
\label{subsec:vckdv}

The solvability condition
\eqref{eq:third_order_solvability_condition} still contains the two
unknown functions $A^{*}$ and $B_1$. They are separated by requiring
the expansion to remain uniformly valid, which leads to a single
evolution equation for $A^{*}$.
We now divide \eqref{eq:third_order_solvability_condition} by $P$ and
use \eqref{eq:ms_characteristic_velocity} and
\eqref{eq:travelling_coordinate}. Since
$\partial/\partial\tau=\partial/\partial\eta$ and
$\partial/\partial x_0=-\partial/\partial\eta$ when acting on $A^{*}$,
the result is
\begin{equation}
\begin{aligned}
\frac{c^2}{c_T^2}
\left(
\frac{\partial^2B_1}{\partial\tau^2}
-\frac{\partial^2B_1}{\partial x_0^2}
\right)
={}&
8\left(\frac{c_T^2}{c_L^2}-1\right)
\frac{\partial^2A^{*}}{\partial\xi\partial\eta}
+4\left(\frac{c_T^2}{c_L^2}-1\right)
\frac{P'}{P}\,
\frac{\partial A^{*}}{\partial\eta}
\\
&+
\frac{4\left(c_L^2-c_T^2\right)\left(c_L^2-2c_T^2\right)^2}{3c_L^6}\,
P^2\,
\frac{\partial^4A^{*}}{\partial\eta^4}
\\
&-
\frac{4}{c_L^6}
\Big[
4c_T^4\eta_1
+3c_L^4\left(\eta_1-\eta_2\right)
+6c_L^2c_T^2\left(\eta_2-\eta_1\right)
\Big]
\frac{\partial A^{*}}{\partial\eta}
\frac{\partial^2A^{*}}{\partial\eta^2}.
\end{aligned}
\label{eq:secularpart_righthandside}
\end{equation}
The right-hand side of \eqref{eq:secularpart_righthandside} depends on
$x_0$ and $\tau$ only through $\eta=\tau-x_0$ and is therefore a
solution of the homogeneous wave equation on the left. Unless it
vanishes, $B_1$ grows linearly with $\tau+x_0$ and the expansion
\eqref{eq:ms_expansion} ceases to be uniformly valid. Setting the
right-hand side equal to zero, dividing by
$8\left(c_T^2/c_L^2-1\right)$ and introducing
\begin{equation}
\phi(\eta,\xi)=\frac{\partial A^{*}}{\partial\eta},
\label{eq:phi_definition}
\end{equation}
we arrive at the variable-coefficient Korteweg--de Vries (KdV) equation
\begin{equation}
\frac{\partial\phi}{\partial\xi}
+\frac{P'}{2P}\,\phi
+\Lambda_1\,\phi\frac{\partial\phi}{\partial\eta}
+\Lambda_2\,P^2\,\frac{\partial^3\phi}{\partial\eta^3}=0,
\label{eq:variable_coefficient_kdv}
\end{equation}
with the constant coefficients
\begin{align}
\Lambda_1
&=
\frac{
4c_T^4\eta_1
+3c_L^4\left(\eta_1-\eta_2\right)
+6c_L^2c_T^2\left(\eta_2-\eta_1\right)
}{2c_L^4\left(c_L^2-c_T^2\right)},
\label{eq:Lambda1_definition}
\\
\Lambda_2
&=
-\frac{\left(c_L^2-2c_T^2\right)^2}{6c_L^4}.
\label{eq:Lambda2_definition}
\end{align}
Because $\partial u/\partial x=-\varepsilon\phi+O(\varepsilon^2)$, the
function $-\varepsilon\phi$ is the leading-order longitudinal strain;
positive values of $\phi$ correspond to compression. Equation
\eqref{eq:variable_coefficient_kdv} describes the evolution of this
strain along the layer, in a frame moving with the velocity $c$. The
surface irregularity enters in two ways: through the linear term
proportional to $P'/P$, and through the factor $P^2$ multiplying the
dispersive term. For $P\equiv1$ the classical KdV equation for the
symmetric long waves in a uniform layer is recovered \cite{Teymur2007}.

The linear geometric term is removed by the transformation
\begin{equation}
\phi(\eta,\xi)=\frac{\psi(\eta,\xi)}{\sqrt{P(\xi)}} ,
\label{eq:phi_psi_transformation}
\end{equation}
which reduces \eqref{eq:variable_coefficient_kdv} to
\begin{equation}
\frac{\partial\psi}{\partial\xi}
+\frac{\Lambda_1}{\sqrt{P}}\,\psi\frac{\partial\psi}{\partial\eta}
+\Lambda_2\,P^2\,\frac{\partial^3\psi}{\partial\eta^3}=0 .
\label{eq:reduced_variable_kdv_xi}
\end{equation}

Before turning to the solutions of
\eqref{eq:reduced_variable_kdv_xi}, we examine its coefficients.
The three coefficients of \eqref{eq:variable_coefficient_kdv} admit
simple mechanical interpretations, which also provide independent
checks of the asymptotic analysis. Let
\begin{equation}
q=1-\frac{2c_T^2}{c_L^2}
=\frac{\lambda}{\lambda+2\mu}
=\frac{\nu}{1-\nu}
\label{eq:q_definition}
\end{equation}
denote the ratio of transverse contraction to longitudinal extension
in a layer with traction-free surfaces under plane strain; by
\eqref{eq:ms_first_order_v_final},
$\partial v_1/\partial y=-q\,\partial u_1/\partial x_0$, and
\eqref{eq:ms_characteristic_velocity} is equivalent to
$c^2=c_L^2\left(1-q^2\right)$.

\medskip
\noindent\emph{Geometric term.}
Multiplying \eqref{eq:variable_coefficient_kdv} by $2P\phi$ and
integrating with respect to $\eta$, for $\phi$ vanishing sufficiently
fast as $|\eta|\rightarrow\infty$, we find
\begin{equation}
\frac{d}{d\xi}\int_{-\infty}^{\infty}P\,\phi^2\,d\eta
=
\frac{d}{d\xi}\int_{-\infty}^{\infty}\psi^2\,d\eta=0 .
\label{eq:energy_flux}
\end{equation}
To leading order, the energy carried by the wave through a
cross-section is proportional to the thickness times the square of
the strain; \eqref{eq:energy_flux} states that this quantity is
conserved, and the term $P'\phi/(2P)$ is precisely what this
conservation requires. A wave entering a thinner part of the layer is
thereby amplified, in proportion to $P^{-1/2}$ as long as nonlinearity
and dispersion are negligible.

\medskip
\noindent\emph{Nonlinear coefficient.}
In terms of $q$, \eqref{eq:Lambda1_definition} takes the form
\begin{equation}
\Lambda_1=\frac{\Gamma}{c^2},
\qquad
\Gamma=\eta_1\left(1-q^3\right)-3\eta_2\,q\left(1-q\right).
\label{eq:Lambda1_compact}
\end{equation}
The quantity $\Gamma$ is the coefficient of the quadratic term in the
uniaxial response of the layer: for a homogeneous deformation
$u_{1,X}=e$ with traction-free surfaces, the stress components of
Appendix~\ref{app:constitutive} give
$T_{11}/\rho_0=c^2e+\Gamma e^2+O(e^3)$. The nonlinear coefficient is
thus independent of the local thickness and involves the third-order
elastic constants only through $\eta_1$ and $\eta_2$.

\medskip
\noindent\emph{Dispersive coefficient.}
Similarly, \eqref{eq:Lambda2_definition} may be written as
\begin{equation}
\Lambda_2
=-\frac{q^2}{6}
=-\frac{1}{6}\left(1-\frac{c^2}{c_L^2}\right)
=-\frac{\nu^2}{6\left(1-\nu\right)^2}.
\label{eq:Lambda2_compact}
\end{equation}
It depends on the linear elastic constants only, is negative for every
material with $\nu\neq0$, and vanishes for $\nu=0$, in which case
dispersion does not appear at this order and the present scaling
ceases to be appropriate. For $P\equiv1$ and infinitesimal amplitudes,
\eqref{eq:variable_coefficient_kdv} admits the plane-wave solutions
$\phi\propto\exp\left[i\left(K\eta+\Lambda_2K^3\xi\right)\right]$,
where $K$ is a dimensionless wavenumber. Returning to the dimensional
variables by means of \eqref{eq:dimensionless_variables} and
\eqref{eq:slow_variable}, we obtain the frequency $\omega=cK/L$ and the
wavenumber $k=K\left(1-\varepsilon\Lambda_2K^2\right)/L$. By
\eqref{eq:epsilon_delta_relation}, $\varepsilon=h^2/L^2$, so that the
correction $\varepsilon K^2$ equals $(kh)^2$ to leading order, $h$
being the half-thickness of the layer. Elimination of $K$ then gives
the dispersion relation
$\omega=ck\left[1-\tfrac{1}{6}q^2(kh)^2\right]$, which coincides with the first
two terms of the long-wave expansion of the fundamental symmetric
branch of the Rayleigh--Lamb frequency equation \cite{EringenSuhubi1975}. In the irregular
layer the dispersive term is multiplied by $P^2$, that is, by the
square of the local thickness.

\section{Soliton-type wave evolution in the variable-coefficient KdV equation}
\label{sec:soliton_evolution}

The reduced equation \eqref{eq:reduced_variable_kdv_xi} is now used to
study the evolution of localized long waves in the irregular layer. It
is first brought to a normalised form in which the dispersive
coefficient is unity and the whole influence of the material and of the
geometry is carried by a single variable coefficient. The sign of this
coefficient, which decides the polarity of the solitary waves, is
discussed next, and its reduction to forms known to be integrable is
then examined.
The formulation of the numerical problem, the numerical method and
its resolution tests, and the numerical results follow.

\subsection{Normalised evolution equation}
\label{subsec:normalised}

By \eqref{eq:Lambda2_compact}, $\Lambda_2<0$ for every material with
$\nu\neq0$. Since $P>0$, the function
\begin{equation}
\theta(\xi)=\left|\Lambda_2\right|\int_0^{\xi}P^2(s)\,ds
\label{eq:new_evolution_variable}
\end{equation}
is strictly increasing and may be used as an evolution variable in
place of $\xi$. Together with $\theta$ we introduce the coordinate
\begin{equation}
\zeta=-\eta=x_0-\tau ,
\label{eq:zeta_definition}
\end{equation}
which measures the distance ahead of the frame moving with the
velocity $c$. Regarding $\psi$ as a function of $(\zeta,\theta)$ and
dividing \eqref{eq:reduced_variable_kdv_xi} by
$\left|\Lambda_2\right|P^2$, we obtain
\begin{equation}
\frac{\partial\psi}{\partial\theta}
+\mathcal{N}(\theta)\,\psi\frac{\partial\psi}{\partial\zeta}
+\frac{\partial^3\psi}{\partial\zeta^3}=0,
\qquad
\mathcal{N}(\theta)
=\mathcal{N}_0\left[P\bigl(\xi(\theta)\bigr)\right]^{-5/2},
\qquad
\mathcal{N}_0=\frac{\Lambda_1}{\Lambda_2}.
\label{eq:normalized_variable_kdv}
\end{equation}
The change of sign of both independent variables leaves the form of
the equation unaltered and ensures that increasing $\theta$
corresponds to increasing distance of propagation. In
\eqref{eq:normalized_variable_kdv} the dispersive coefficient is
unity; the geometry enters through the factor $P^{-5/2}$ of the
nonlinear coefficient, and the material through the single number
$\mathcal{N}_0$. This normalisation is preferred to the equivalent one
in which the nonlinear coefficient is reduced to a constant, because
$\Lambda_2$ has a fixed sign and does not vanish, whereas $\Lambda_1$
may have either sign and may vanish, as shown below.

Two consequences of \eqref{eq:normalized_variable_kdv} are noted.
First, in the normalised variables infinitesimal disturbances satisfy
the linearised KdV equation of the uniform layer whatever the profile
$P$; the variation of the thickness then acts only through the
amplitude factor in \eqref{eq:phi_psi_transformation} and through the
stretching \eqref{eq:new_evolution_variable}, and any further effect of
the geometry on $\psi$ is a nonlinear one. Secondly, the substitution
$\psi\rightarrow6\psi/\mathcal{N}_0$ replaces $\mathcal{N}_0$ by $6$,
so that different materials are distinguished, in the normalised
variables, only by the scale and the sign of the amplitude.

Since $\mathcal{N}$ does not depend on $\zeta$, for solutions decaying
sufficiently fast as $|\zeta|\rightarrow\infty$, or periodic in
$\zeta$, equation \eqref{eq:normalized_variable_kdv} possesses the
invariants
\begin{equation}
I_1=\int\psi\,d\zeta,
\qquad
I_2=\int\psi^2\,d\zeta ,
\label{eq:invariants}
\end{equation}
the second of which is the energy integral \eqref{eq:energy_flux}.

\subsection{Sign of the nonlinear coefficient and polarity of solitary waves}
\label{sec:nonlinear_coefficient_sign}

The nonlinear coefficient of \eqref{eq:normalized_variable_kdv} keeps
the sign of $\mathcal{N}_0$ along the whole layer: the irregularity
changes the magnitude of $\mathcal{N}$, but cannot reverse its sign as
long as $P>0$. From \eqref{eq:Lambda1_compact} and
\eqref{eq:Lambda2_compact},
\begin{equation}
\mathcal{N}_0
=-\frac{6\,\Gamma}{q^2c^2}
=-\frac{3\,\mathcal{F}}
{\left(c_L^2-c_T^2\right)\left(c_L^2-2c_T^2\right)^2},
\qquad
\mathcal{F}
=\left(3c_L^4-6c_L^2c_T^2+4c_T^4\right)\eta_1
-3c_L^2\left(c_L^2-2c_T^2\right)\eta_2 ,
\label{eq:sign_explicit_nonlinearity}
\end{equation}
so that
$\operatorname{sgn}\mathcal{N}_0=-\operatorname{sgn}\Gamma
=-\operatorname{sgn}\mathcal{F}$. The numerator involves the
third-order elastic constants, whereas the denominator involves the
linear constants only. The coefficient vanishes when
\begin{equation}
\eta_2=\eta_{2,\mathrm{cr}}
=\frac{3c_L^4-6c_L^2c_T^2+4c_T^4}
{3c_L^2\left(c_L^2-2c_T^2\right)}\,\eta_1 ,
\label{eq:sign_critical_material_parameter}
\end{equation}
and, for $c_L^2>2c_T^2$, that is, for $\nu>0$, we have
$\mathcal{N}_0>0$ if $\eta_2>\eta_{2,\mathrm{cr}}$ and
$\mathcal{N}_0<0$ if $\eta_2<\eta_{2,\mathrm{cr}}$. At
$\eta_2=\eta_{2,\mathrm{cr}}$ the quadratic nonlinearity disappears,
\eqref{eq:normalized_variable_kdv} becomes linear, and the description
of nonlinear waves requires a different scaling in which higher-order
terms are retained; this case is not considered here.

For a layer of uniform thickness, $\mathcal{N}\equiv\mathcal{N}_0$,
and \eqref{eq:normalized_variable_kdv} has the solitary-wave solution \cite{Remoissenet2013}
\begin{equation}
\psi(\zeta,\theta)
=\frac{3c_s}{\mathcal{N}_0}\,
\operatorname{sech}^{2}\!\left[
\frac{\sqrt{c_s}}{2}\left(\zeta-\zeta_0-c_s\theta\right)
\right],
\qquad c_s>0,
\label{eq:sign_constant_coefficient_soliton}
\end{equation}
where $\zeta_0$ is the initial position of the crest and $c_s$ is the
velocity in the $(\zeta,\theta)$-plane. The solitary wave advances
towards positive $\zeta$, that is, it travels faster than $c$, and its
amplitude has the sign of $\mathcal{N}_0$. Recalling that positive
$\phi$ means compression, we conclude that a material with $\Gamma<0$,
which by \eqref{eq:Lambda1_compact} is stiffer in compression than in
extension, supports compressive solitary waves, whereas a material
with $\Gamma>0$ supports tensile ones. A pulse whose polarity is
opposite to that of \eqref{eq:sign_constant_coefficient_soliton} does
not evolve into solitary waves but disperses into an oscillatory wave
train \cite{Remoissenet2013}. When $P$ varies, \eqref{eq:sign_constant_coefficient_soliton} is
no longer an exact solution; it will serve as the initial condition
and as the reference state in the numerical experiments.

\subsection{Integrable cases}
\label{subsec:integrability}

It is natural to ask whether, for some profiles of the free surfaces,
\eqref{eq:normalized_variable_kdv} reduces to an equation known to be
integrable. For equations of the form
$\psi_\theta+f(\theta)\psi\psi_\zeta+g(\theta)\psi_{\zeta\zeta\zeta}=0$,
the Painlev\'e analysis leads to the integrability condition
$g=f\left(\alpha\int f\,d\theta+\beta\right)$, with constants $\alpha$
and $\beta$ \cite{Zhang2006,Tian2006,TianZhang2014}. For $g=1$ and
$f=\mathcal{N}$ this condition is satisfied when $\mathcal{N}$ is constant, which is the uniform layer, and
$\mathcal{N}\propto\left(\theta+\theta_0\right)^{-1/2}$ with a
constant $\theta_0$. The second case is made explicit by the
substitution
\begin{equation}
w=\frac{\mathcal{N}_0}{6}\,P^{-5/2}\,\psi ,
\label{eq:w_definition}
\end{equation}
which transforms \eqref{eq:normalized_variable_kdv} into
\begin{equation}
\frac{\partial w}{\partial\theta}
+\frac{5}{2P}\frac{dP}{d\theta}\,w
+6w\frac{\partial w}{\partial\zeta}
+\frac{\partial^3w}{\partial\zeta^3}=0 .
\label{eq:w_equation}
\end{equation}
This is the cylindrical KdV equation \cite{Tian2006,DaiHuo2002,Ablowitz2016}, with the linear term
$w/[2(\theta+\theta_0)]$, if and only if
$P^5\propto\theta+\theta_0$. Normalising $P=1$ at $\xi=0$ and
returning to the physical variable by means of
\eqref{eq:new_evolution_variable}, we find
\begin{equation}
P(\xi)=\left(1+\frac{\xi}{\xi_0}\right)^{1/3},
\qquad
\xi_0=\frac{5\,\theta_0}{3\left|\Lambda_2\right|}.
\label{eq:integrable_profile}
\end{equation}
Hence, for a layer of uniform thickness and for the monotone profile
\eqref{eq:integrable_profile}, \eqref{eq:normalized_variable_kdv}
reduces to the KdV and to the cylindrical KdV equation, respectively,
both of which are known to be integrable. In the second case the
thickness increases
($\xi_0>0$) or decreases ($\xi_0<0$) as the cube root of the distance.
For $|\xi|\ll|\xi_0|$ this profile coincides, to first order, with a
linear taper of small slope.

The profile \eqref{eq:integrable_profile} is, however, admissible in
a bounded region only. For $\xi_0>0$ the thickness grows without
bound, and the long-wave assumption underlying the present analysis,
which requires $\varepsilon P^2\ll1$, fails at distances
$\xi/\xi_0=O(\varepsilon^{-3/2})$; for $\xi_0<0$ the thickness
vanishes at $\xi=|\xi_0|$. For this reason the cylindrical case is not
examined further. On the other hand, bounded non-constant profiles, and in particular
periodic or localized irregularities, do not satisfy the above
condition, so that \eqref{eq:normalized_variable_kdv} does not reduce
to either of these integrable forms. Its solutions are therefore
obtained numerically in what follows.

\subsection{Formulation of the numerical problem}
\label{subsec:numerical_formulation}

For the numerical computations we write
\begin{equation}
x=\zeta,\qquad t=\theta,\qquad u(x,t)=\psi(\zeta,\theta),
\label{eq:numerical_notation}
\end{equation}
so that, throughout this section, $x$ and $t$ denote the normalised
coordinate and the normalised propagation distance, and $u$ denotes
the normalised strain amplitude $\psi$; they are not related to the
physical coordinate and time of Section~\ref{sec:formulation} or to
the dimensionless displacement component $u$ of
Section~\ref{subsec:dimensionless}.
Equation \eqref{eq:normalized_variable_kdv} then reads
\begin{equation}
\frac{\partial u}{\partial t}
+\mathcal{N}(t)\,u\frac{\partial u}{\partial x}
+\frac{\partial^3u}{\partial x^3}=0,
\qquad
\mathcal{N}(t)=\mathcal{N}_0\left[\mathcal{P}(t)\right]^{-5/2},
\qquad
\mathcal{P}(t)=P\bigl(\xi(t)\bigr).
\label{eq:num_kdv_reference}
\end{equation}
The geometry may be specified either by $P(\xi)$, in which case $t$
follows from \eqref{eq:new_evolution_variable}, or directly by the
composed function $\mathcal{P}(t)$, in which case the physical
profile is recovered from
\begin{equation}
\xi(t)=\frac{1}{\left|\Lambda_2\right|}
\int_0^{t}\frac{dt'}{\mathcal{P}^2(t')} .
\label{eq:xi_of_t}
\end{equation}
The second alternative is adopted here, with the periodic modulation
\begin{equation}
\mathcal{P}(t)=1+A\cos(\omega t),
\qquad 0\leq A<1 .
\label{eq:num_reference_modulation}
\end{equation}
By \eqref{eq:xi_of_t}, this corresponds to a surface profile that is
periodic in $\xi$, with the same extreme values $1\pm A$ of the
half-thickness and with the period
\begin{equation}
\Xi=\frac{2\pi}{\omega\left|\Lambda_2\right|\left(1-A^2\right)^{3/2}} .
\label{eq:physical_period}
\end{equation}
The nonlinear coefficient accordingly oscillates between
$\mathcal{N}_0(1+A)^{-5/2}$ and $\mathcal{N}_0(1-A)^{-5/2}$.

The values of $\mathcal{N}_0$ used in the computations are collected
in Table~\ref{tab:materials}. For iron and copper, the second- and
third-order elastic constants are taken directly from Truesdell and
Noll \cite[Sect.~66]{TruesdellNoll1992}.
Because $\mathcal{N}_0$ depends on the material only through ratios
of elastic constants, the mass density is not required. In addition to
the two metals, model
materials with $c_L^2/c_T^2=3$ $(\nu=1/4)$ are considered, for which
$\mathcal{N}_0=3\left(9\eta_2-13\eta_1\right)/(2c_T^2)$; the ratio
$\eta_2/\eta_1$ is chosen so as to give prescribed values of
$\mathcal{N}_0$ of either sign.

\par\vspace{6mm}
\noindent
\begin{minipage}{\linewidth}
\centering
\captionsetup{type=table,skip=8pt}
\captionof{table}{Material parameters of the normalised equation
\eqref{eq:num_kdv_reference}. For the model materials
$c_L^2/c_T^2=3$ and $\eta_1/c_T^2=1$.}
\label{tab:materials}
\begin{tabular}{lcccc}
\toprule
Material & $\nu$ & $\Lambda_1$ & $\Lambda_2$ & $\mathcal{N}_0$ \\
\midrule
Iron   & $0.291$ & $-4.804$ & $-0.02805$ & $171.29$ \\
Copper & $0.346$ & $-5.856$ & $-0.04666$ & $125.49$ \\
\midrule
Model, $\eta_2/\eta_1=17/9$  & $1/4$ & $-1/9$  & $-1/54$ & $6$  \\
Model, $\eta_2/\eta_1=43/27$ & $1/4$ & $-1/27$ & $-1/54$ & $2$  \\
Model, $\eta_2/\eta_1=29/9$  & $1/4$ & $-4/9$  & $-1/54$ & $24$ \\
Model, $\eta_2/\eta_1=1$     & $1/4$ & $1/9$   & $-1/54$ & $-6$ \\
\bottomrule
\end{tabular}
\end{minipage}
\par\vspace{6mm}

For both metals $\mathcal{N}_0>0$, that is, $\Gamma<0$: iron and
copper support compressive solitary waves, and their normalised
equations differ only in the scale of the amplitude.

\subsection{Numerical method and resolution tests}
\label{subsec:numerical_method}

In all computations the initial condition is the pulse
\begin{equation}
u(x,0)=\frac{c_s}{2}\,
\operatorname{sech}^{2}\!\left(\frac{\sqrt{c_s}}{2}\,x\right),
\label{eq:num_reference_initial}
\end{equation}
which has the amplitude $c_s/2$ and the characteristic width
$2/\sqrt{c_s}$. By \eqref{eq:sign_constant_coefficient_soliton}, it
is an exact solitary wave of the uniform layer when
$\mathcal{N}_0=6$; for other values of $\mathcal{N}_0$ it is not, and
its evolution in the uniform layer $(A=0)$ then provides the baseline
against which the effect of the irregularity is assessed.

Equation \eqref{eq:num_kdv_reference} is solved on a periodic interval
$-R\leq x<R$ by a Fourier pseudospectral method with $N$ grid points.
The dispersive term is integrated exactly through its exponential
propagator and the nonlinear term is advanced by the fourth-order
exponential time-differencing Runge--Kutta scheme (ETDRK4)
\cite{CoxMatthews2002,KassamTrefethen2005}.
The coefficient $\mathcal{N}(t)$ is evaluated at the beginning of each
step and kept fixed within the step. Unless stated otherwise,
\begin{equation}
R=400,\qquad N=1024,\qquad \Delta t=0.01,\qquad t_{\max}=100,
\qquad \omega=0.2 ,
\label{eq:num_reference_resolution}
\end{equation}
so that $\Delta x=0.78125$ and the interval $0\leq t\leq100$ contains
about three periods $2\pi/\omega\simeq31.42$ of the modulation
\eqref{eq:num_reference_modulation}.

The resolution was examined for the reference case defined below
($\mathcal{N}_0=6$, $A=0.01$, $c_s=0.1$) by means of the relative
differences
\begin{equation}
E_2=\frac{\left\|u_{\mathrm{test}}-u_{\mathrm{ref}}\right\|_2}
{\left\|u_{\mathrm{ref}}\right\|_2},
\qquad
E_\infty=\frac{\left\|u_{\mathrm{test}}-u_{\mathrm{ref}}\right\|_\infty}
{\left\|u_{\mathrm{ref}}\right\|_\infty}
\label{eq:kdv_relative_profile_errors}
\end{equation}
between profiles at $t=100$, evaluated at common grid points without
any translation. In the grid test $\Delta t=0.01$ is fixed and the
solution with $N=4096$ is the reference; in the step test $N=1024$ is
fixed and the solution with $\Delta t=0.0025$ is the reference. The
results are given in Table~\ref{tab:kdv_refinement}. The spatial error
is at the level of round-off already for $N=1024$. The differences in
the step test decrease in proportion to
$\Delta t-\Delta t_{\mathrm{ref}}$, as expected when the coefficient is
frozen within a step; at the adopted step $\Delta t=0.01$ they are
$E_2=2.5\times10^{-6}$ and $E_\infty=1.5\times10^{-6}$, which is far
below the differences between the cases compared in what follows. The
temporal discretization is therefore the dominant, and still
negligible, source of numerical error.

\par\vspace{6mm}
\noindent
\begin{minipage}{\linewidth}
\centering
\captionsetup{type=table,skip=8pt}
\captionof{table}{Grid and step refinement for the reference case at $t=100$.}
\label{tab:kdv_refinement}
\begin{tabular}{llcc}
\toprule
Test & Resolution & $E_2$ & $E_\infty$ \\
\midrule
Grid & $N=512$  & $1.48\times10^{-8}$  & $1.02\times10^{-8}$ \\
Grid & $N=1024$ & $6.3\times10^{-16}$  & $4.2\times10^{-16}$ \\
Grid & $N=2048$ & $6.4\times10^{-16}$  & $5.6\times10^{-16}$ \\
Step & $\Delta t=0.04$  & $1.2515\times10^{-5}$ & $7.3895\times10^{-6}$ \\
Step & $\Delta t=0.02$  & $5.8413\times10^{-6}$ & $3.4477\times10^{-6}$ \\
Step & $\Delta t=0.01$  & $2.5036\times10^{-6}$ & $1.4774\times10^{-6}$ \\
Step & $\Delta t=0.005$ & $8.3458\times10^{-7}$ & $4.9245\times10^{-7}$ \\
\bottomrule
\end{tabular}
\end{minipage}
\par\vspace{6mm}

As a further check, the invariant $I_2$ of \eqref{eq:invariants} is
monitored in every computation. Its conservation tests the consistency
of the discretization but is not, by itself, a measure of the error of
the profile, which is why it is used together with
Table~\ref{tab:kdv_refinement}.

\subsection{Numerical results}
\label{sec:numerical_evaluations}

\noindent\emph{Reference computation.}
We begin with the model material for which $\mathcal{N}_0=6$
$(\eta_2/\eta_1=17/9)$, a weak irregularity with $A=0.01$, and the
pulse \eqref{eq:num_reference_initial} with $c_s=0.1$, whose amplitude
is $0.05$ and whose width is $2/\sqrt{c_s}\simeq6.32$. In this case
\begin{equation}
\mathcal{N}(t)=6\left[1+0.01\cos(0.2\,t)\right]^{-5/2},
\qquad
5.853\leq\mathcal{N}(t)\leq6.153 ,
\label{eq:num_reference_coefficient}
\end{equation}
so that the coefficient departs from the value for which the pulse is
an exact solitary wave by abo $2.5\%$. The results are shown
in Fig.~\ref{fig:kdv_reference}. The pulse advances towards positive
$x$ and retains its shape; at $t=100$ its maximum is $0.049937$, which
is $0.13\%$ below the initial value, and it is attained at the grid
point $x=10.156$, close to the displacement $c_st=10$ of the solitary
wave of the uniform layer. No secondary structure is visible on the
scale of the figure. The deviation of $I_2$ from its initial value
does not exceed $1.5\times10^{-15}$ in absolute and
$7.3\times10^{-14}$ in relative terms, and is indistinguishable from
zero in Fig.~\ref{fig:kdv_reference}(b). A weak periodic irregularity
thus leaves the solitary wave practically unaffected over the
distance considered, and this computation serves as the baseline for
those that follow.

\par\vspace{6mm}
\noindent
\begin{minipage}{\linewidth}
\centering
\captionsetup{skip=8pt}
\includegraphics[width=\linewidth]{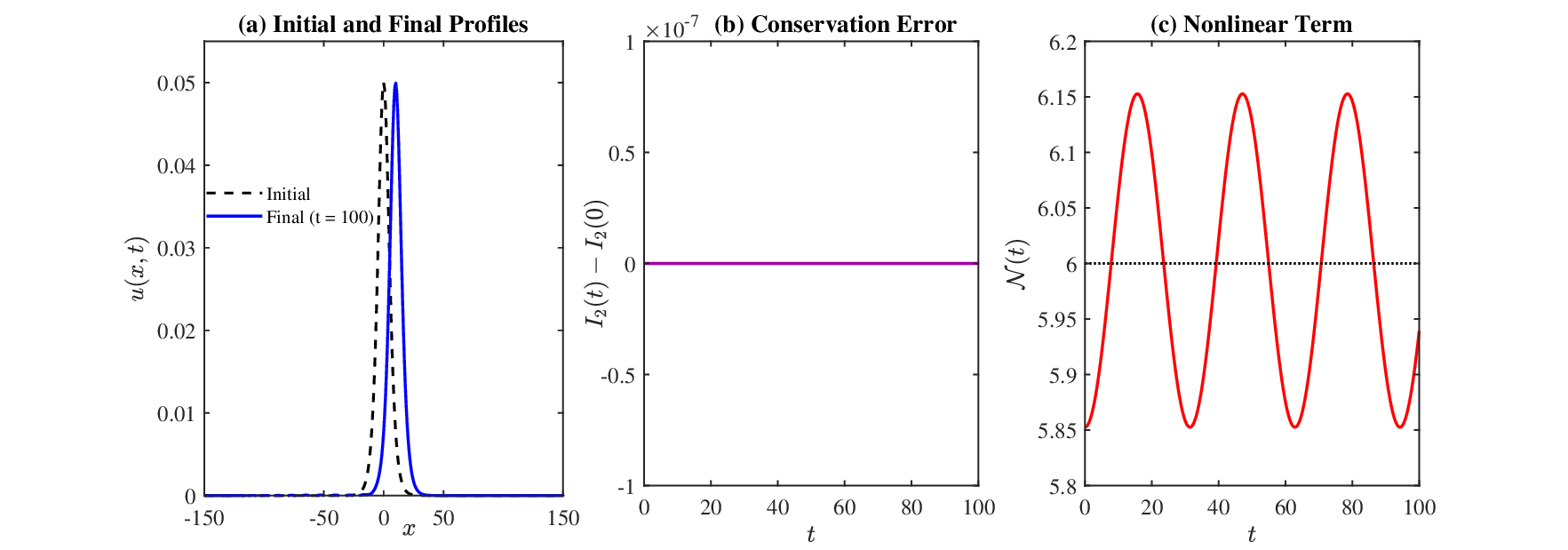}
\captionof{figure}{Reference computation for the model material with
$\mathcal{N}_0=6$ $(c_L^2/c_T^2=3$, $\eta_2/\eta_1=17/9)$ and the
irregularity $A=0.01$, $\omega=0.2$: (a) initial profile and profile
at $t=100$ for $c_s=0.1$; (b) deviation of the invariant $I_2$ from
its initial value, on a fixed scale of $\pm10^{-7}$; (c) nonlinear
coefficient $\mathcal{N}(t)$, the dotted line marking
$\mathcal{N}=6$.}
\label{fig:kdv_reference}
\end{minipage}
\par\vspace{6mm}

\medskip
\noindent\emph{Iron.}
The effect of the amplitude of the irregularity is examined next for
iron, for which $\mathcal{N}_0=171.29$
(Table~\ref{tab:materials}). The amplitude takes the values $A=0$,
$0.01$, $0.05$ and $0.10$, the case $A=0$ being the layer of uniform
thickness, and
\begin{equation}
\mathcal{N}(t)=171.29\left[1+A\cos(0.2\,t)\right]^{-5/2}
\label{eq:iron_nonlinear_coefficient}
\end{equation}
varies over the intervals $[167.08,\,175.64]$, $[151.62,\,194.72]$ and
$[134.97,\,222.90]$ for $A=0.01$, $0.05$ and $0.10$, respectively.
Since the geometric factor enters through an inverse power, the
variation is not symmetric about $\mathcal{N}_0$,
\begin{equation}
\frac{\mathcal{N}(t)}{\mathcal{N}_0}
=1-\frac{5}{2}A\cos(\omega t)
+\frac{35}{8}A^2\cos^2(\omega t)+O(A^3),
\label{eq:roughness_small_A_expansion}
\end{equation}
and its mean over a period exceeds $\mathcal{N}_0$ by a term of order
$A^2$. The initial pulse is \eqref{eq:num_reference_initial} with
$c_s=0.01$, that is, with the amplitude $0.005$ and the width $20$;
the same pulse is used for the four values of $A$. Because
$\mathcal{N}_0>0$, this pulse has the polarity of the solitary waves of
the material, but it is much larger than the solitary wave of the
same width, whose amplitude is $3c_s/\mathcal{N}_0=1.75\times10^{-4}$.

Figure~\ref{fig:iron_roughness} shows that in all four cases the
pulse steepens on its leading side and disintegrates into a train of
solitary waves ordered by amplitude, the largest one in front. For
the uniform layer, the first three maxima at $t=100$ are
$8.623\times10^{-3}$ at $x=60.94$, $6.187\times10^{-3}$ at $x=42.97$
and $3.801\times10^{-3}$ at $x=26.56$, while the rear part of the
pulse has not yet separated. No negative values of any significance
occur. This is the classical disintegration of an initial pulse into a train
of solitary waves ordered by amplitude \cite{Remoissenet2013}.

The irregularity does not alter this picture qualitatively, but it
changes the individual waves measurably. As $A$ increases through
$0.01$, $0.05$ and $0.10$, the leading maximum decreases to
$8.519\times10^{-3}$, $8.360\times10^{-3}$ and $8.065\times10^{-3}$,
the last value being $6.5\%$ below that of the uniform layer, whereas
the second and third maxima increase, to $6.366\times10^{-3}$ and
$4.038\times10^{-3}$ for $A=0.10$, that is, by $2.9\%$ and $6.2\%$.
The positions of the maxima change by at most one grid spacing. For
$A=0.10$ additional undulations appear in the rear part of the pulse,
and a shallow depression of depth $2.3\times10^{-4}$ forms behind the
leading wave. The irregularity thus redistributes the amplitudes
within the train rather than amplifying or attenuating it as a whole.
The overall difference is measured by
\begin{equation}
E_A=\frac{\left\|u_A(\cdot,100)-u_{0}(\cdot,100)\right\|_2}
{\left\|u_{0}(\cdot,100)\right\|_2},
\label{eq:iron_profile_difference}
\end{equation}
where $u_A$ denotes the solution for the amplitude $A$; we find
$E_{0.01}=2.1\%$, $E_{0.05}=7.0\%$ and $E_{0.10}=16.8\%$. These
differences are several orders of magnitude larger than the numerical
errors of Table~\ref{tab:kdv_refinement}. In these computations the
relative deviation of $I_2$ remains below $7\times10^{-10}$, and the
solution in the outer region $|x|\geq360$ never exceeds
$9\times10^{-9}$ of its overall maximum, so that the periodic boundary
conditions have no influence on the results. Doubling $N$ and halving
$\Delta t$ for $A=0.10$ change the final profile by $4\times10^{-7}$
and $5\times10^{-5}$ in the sense of $E_2$, respectively.

\par\vspace{6mm}
\noindent
\begin{minipage}{\linewidth}
\centering
\captionsetup{skip=8pt}
\includegraphics[width=\linewidth]{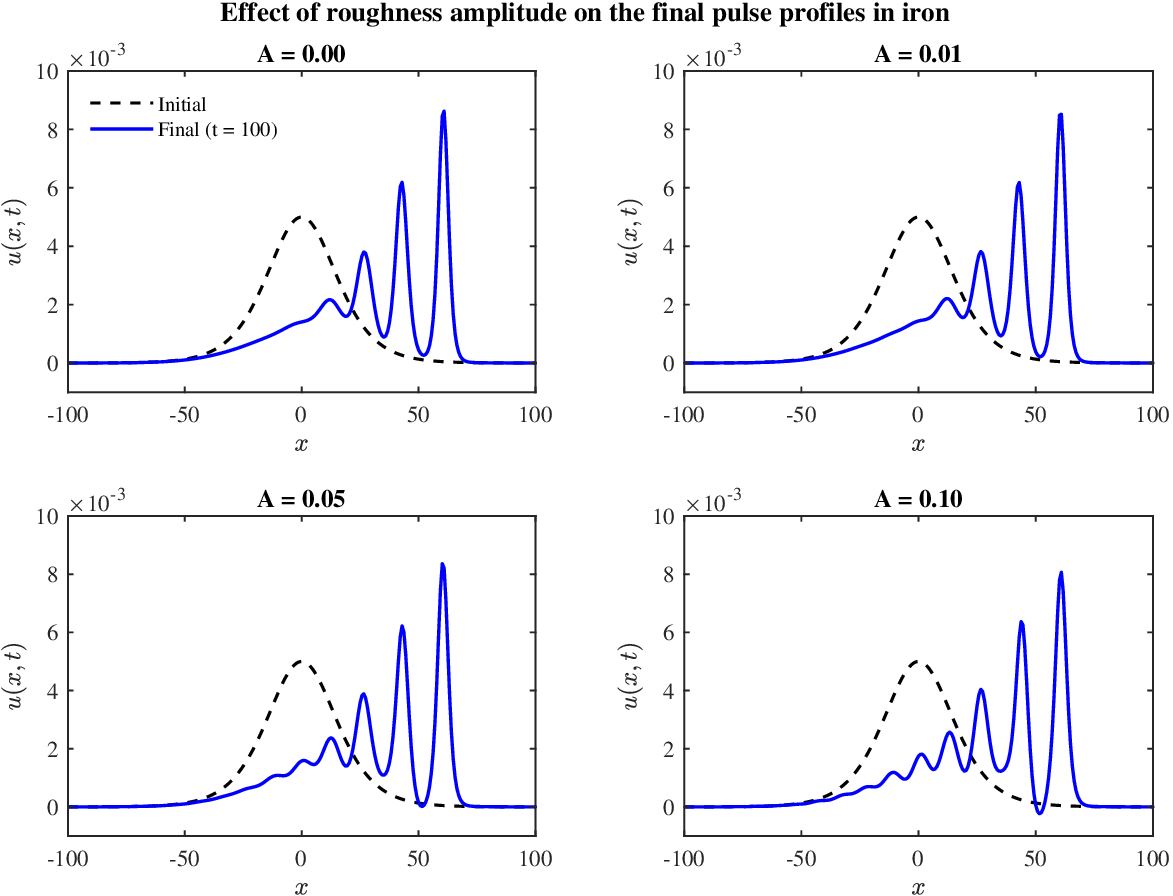}
\captionof{figure}{Effect of the amplitude of the irregularity on the evolution
of a pulse in iron ($\mathcal{N}_0=171.29$), for $A=0$, $0.01$, $0.05$
and $0.10$ and $\omega=0.2$. Dashed curves: common initial profile
with $c_s=0.01$; solid curves: solutions at $t=100$. The axis limits
are the same in all panels.}
\label{fig:iron_roughness}
\end{minipage}
\par\vspace{6mm}

\medskip
\noindent\emph{Copper.}
The same comparison is made for copper, for which
$\mathcal{N}_0=125.49$, with the initial pulse, the frequency and the
numerical parameters of the iron computations. The coefficient now
varies over $[122.41,\,128.68]$, $[111.08,\,142.66]$ and
$[98.88,\,163.31]$ for $A=0.01$, $0.05$ and $0.10$. For a given $A$
the relative modulation $\mathcal{N}(t)/\mathcal{N}_0$ is the same for
the two metals; they differ in the strength of the nonlinearity
relative to the dispersion, the value for copper being $73\%$ of that
for iron. It should be kept in mind that equal values of $t$ do not
correspond to equal distances in the two materials, because the
relation \eqref{eq:new_evolution_variable} between $t$ and $\xi$
involves $\Lambda_2$.

The evolution, shown in Fig.~\ref{fig:copper_roughness}, is of the
same type as in iron, but it proceeds more slowly. For the uniform
layer the first three maxima are $8.245\times10^{-3}$ at $x=46.09$,
$5.004\times10^{-3}$ at $x=29.69$ and $2.834\times10^{-3}$ at
$x=14.06$; the leading wave has travelled a shorter distance than in
iron, and fewer waves have emerged. With increasing $A$ the
leading maximum decreases to $8.196\times10^{-3}$,
$8.044\times10^{-3}$ and $7.937\times10^{-3}$, the reduction for
$A=0.10$ being $3.7\%$, while the second and third maxima increase to
$5.246\times10^{-3}$ and $3.038\times10^{-3}$, that is, by $4.8\%$ and
$7.2\%$. The profile differences \eqref{eq:iron_profile_difference},
referred to the uniform copper layer, are $1.3\%$, $4.3\%$ and
$9.3\%$, compared with $2.1\%$, $7.0\%$ and $16.8\%$ for iron. Under
identical normalised conditions copper is therefore less sensitive to
the irregularity than iron, in accordance with its smaller nonlinear
coefficient. The relative deviation of $I_2$ is below
$8\times10^{-13}$ and the solution in $|x|\geq360$ is below
$1.1\times10^{-11}$ of its maximum.

\par\vspace{6mm}
\noindent
\begin{minipage}{\linewidth}
\centering
\captionsetup{skip=8pt}
\includegraphics[width=\linewidth]{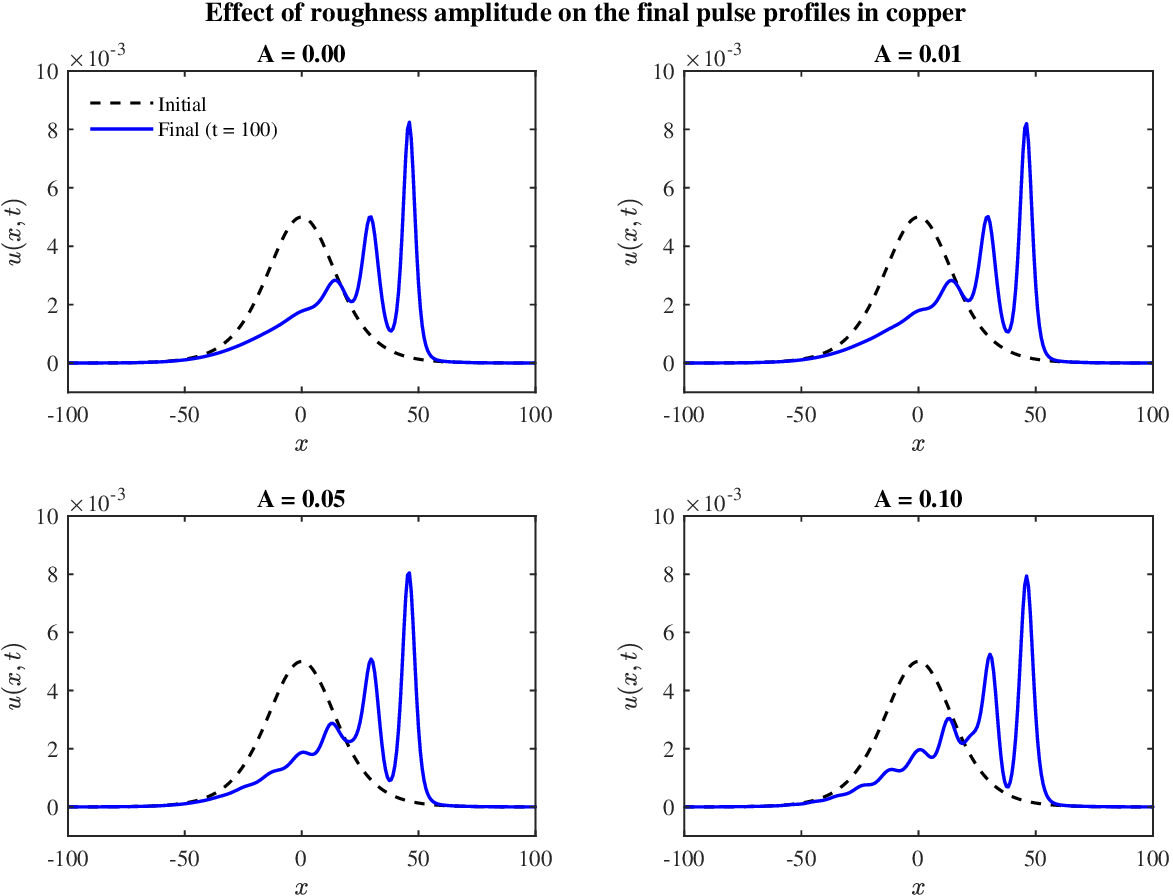}
\captionof{figure}{Effect of the amplitude of the irregularity on the evolution
of a pulse in copper ($\mathcal{N}_0=125.49$), for $A=0$, $0.01$,
$0.05$ and $0.10$ and $\omega=0.2$. Dashed curves: common initial
profile with $c_s=0.01$; solid curves: solutions at $t=100$. The axis
limits are the same in all panels.}
\label{fig:copper_roughness}
\end{minipage}
\par\vspace{6mm}

\medskip
\noindent\emph{Velocity of the leading wave.}
To quantify the influence of the irregularity on the motion of the
leading wave, the iron computations are repeated for $A=0$, $0.02$,
\ldots, $0.10$, and the position $x_{\mathrm p}(t;A)$ of the largest
maximum is determined. Since the maximum falls on the same or on a
neighbouring grid point in all cases, its position is refined by a
parabola through the largest grid value $u_j$ and its two neighbours,
\begin{equation}
x_{\mathrm p}\simeq x_j+\frac{\Delta x}{2}\,
\frac{u_{j-1}-u_{j+1}}{u_{j-1}-2u_j+u_{j+1}} .
\label{eq:peak_parabolic_interpolation}
\end{equation}
Table~\ref{tab:iron_peak_speed} gives $x_{\mathrm p}$ at $T=100$, the
mean velocity $\overline{v}_{\mathrm p}=x_{\mathrm p}(T;A)/T$ and its
change relative to the uniform layer,
$\delta\overline{v}_{\mathrm p}
=100\,[\overline{v}_{\mathrm p}(A)/\overline{v}_{\mathrm p}(0)-1]$.
The mean velocity depends on $A$ weakly and not monotonically: it is
smallest for $A=0.04$, where it is $0.37\%$ below the value for the
uniform layer, and exceeds that value by $0.19\%$ for $A=0.10$. These
figures refer to the normalised variables and to an interval that is
not an integer multiple of the period of the modulation; they show
that the irregularity affects the velocity of the leading wave much
less than its amplitude.

\par\vspace{6mm}
\noindent
\begin{minipage}{\linewidth}
\centering
\captionsetup{type=table,skip=8pt}
\captionof{table}{Position and mean velocity of the leading maximum in iron at
$T=100$.}
\label{tab:iron_peak_speed}
\begin{tabular}{cccc}
\toprule
$A$ & $x_{\mathrm p}(100;A)$ & $\overline{v}_{\mathrm p}(A)$ &
$\delta\overline{v}_{\mathrm p}$ (\%) \\
\midrule
$0.00$ & $60.6563$ & $0.60656$ & $0.000$ \\
$0.02$ & $60.4987$ & $0.60499$ & $-0.260$ \\
$0.04$ & $60.4335$ & $0.60433$ & $-0.367$ \\
$0.06$ & $60.4572$ & $0.60457$ & $-0.328$ \\
$0.08$ & $60.5732$ & $0.60573$ & $-0.137$ \\
$0.10$ & $60.7721$ & $0.60772$ & $0.191$ \\
\bottomrule
\end{tabular}
\end{minipage}
\par\vspace{6mm}

\medskip
\noindent\emph{Model materials.}
Finally, in order to separate the role of the material nonlinearity
from that of the geometry, we consider the four model materials of
Table~\ref{tab:materials}, for which $\mathcal{N}_0=6$, $2$, $24$ and
$-6$, with the same irregularity, $A=0.05$ and $\omega=0.2$, and the
same initial pulse, $c_s=0.1$. Here
$\mathcal{N}(t)=\mathcal{N}_0[1+0.05\cos(0.2\,t)]^{-5/2}$ varies
between $0.885\,\mathcal{N}_0$ and $1.137\,\mathcal{N}_0$. The
computations use $R=800$, $N=4096$ and $\Delta t=0.005$; the larger
interval accommodates the oscillatory tails. The results are shown in
Fig.~\ref{fig:artificial_material_response}.

For $\mathcal{N}_0=6$ (panel a) the pulse remains a solitary wave: its
maximum at $t=100$ is $0.04992$, which is $0.16\%$ below the initial
value, and only a tail of very small amplitude is formed. For
$\mathcal{N}_0=2$ (panel b) the nonlinearity is too weak to balance
the dispersion of the given pulse; the maximum falls to $0.04136$,
that is, by $17.3\%$, and an oscillatory tail with a first depression
of $-0.0092$ develops behind it. For $\mathcal{N}_0=24$ (panel c) the
pulse is larger than the solitary wave of its width and disintegrates:
at $t=100$ two separated waves with the maxima $0.06734$ at $x=57.8$
and $0.02470$ at $x=18.0$ are present, the leading one being $34.7\%$
higher than the initial pulse. For the uniform layer the inverse
scattering theory gives in this case three solitary waves with the
amplitudes $0.0703$, $0.0235$ and $0.0017$, and one wave with the
amplitude $0.0314$ for $\mathcal{N}_0=2$. For $\mathcal{N}_0=-6$
(panel d) the polarity of the pulse is opposite to that of the
solitary waves of the material; no solitary wave is formed, and the
pulse disperses into an extended oscillatory wave train, its maximum
decreasing to $0.02883$ and the deepest depression reaching
$-0.02297$. In these computations the relative deviation of $I_2$
does not exceed $2.6\times10^{-12}$, and repeating them on the
interval with $R=400$ and $N=2048$ changes the profiles in
$-200\leq x\leq100$ by less than $5.4\times10^{-7}$.

\par\vspace{6mm}
\noindent
\begin{minipage}{\linewidth}
\centering
\captionsetup{skip=8pt}
\includegraphics[width=\linewidth]{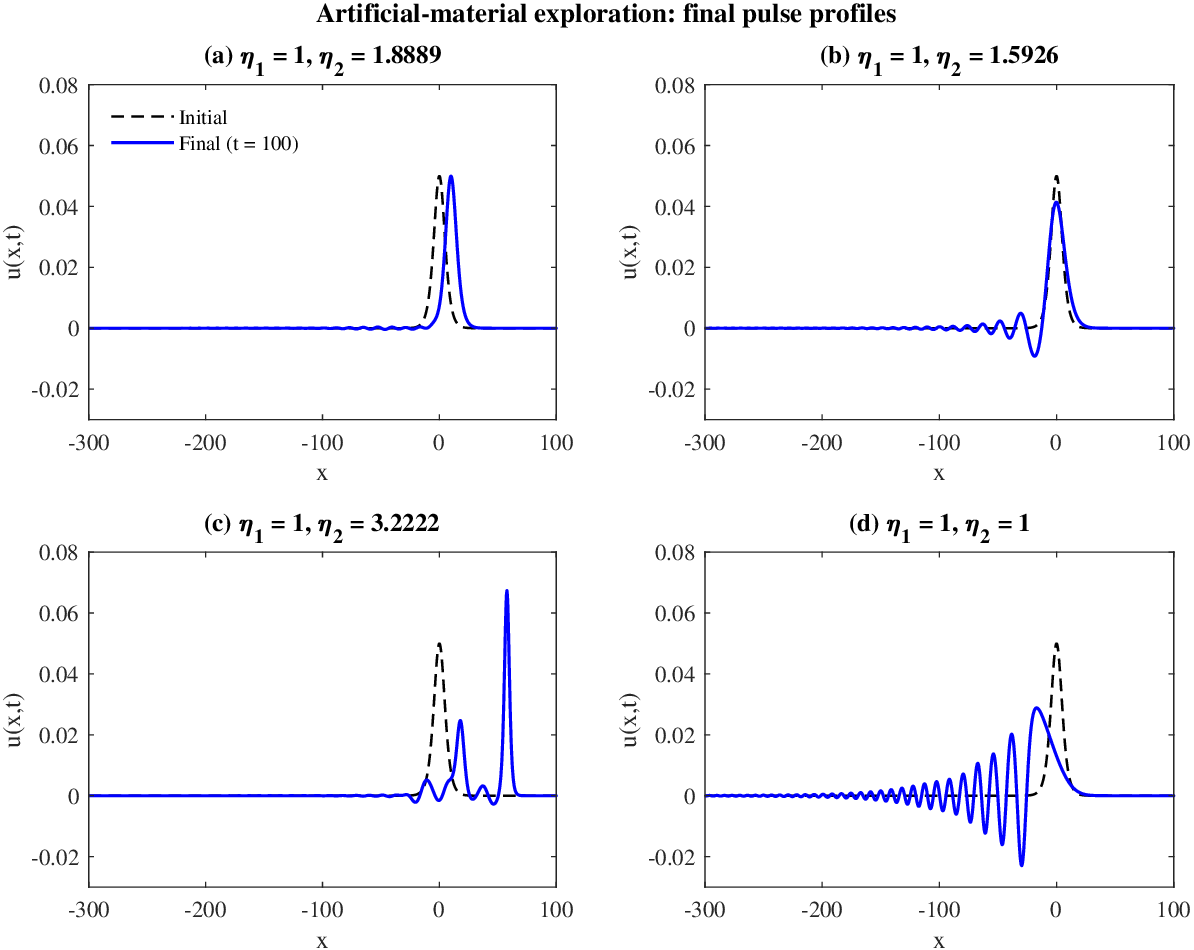}
\captionof{figure}{Evolution of the same pulse $(c_s=0.1)$ in four model
materials with $c_L^2/c_T^2=3$ and $\eta_1/c_T^2=1$:
(a) $\eta_2/\eta_1=17/9$, (b) $43/27$, (c) $29/9$ and (d) $1$,
corresponding to $\mathcal{N}_0=6$, $2$, $24$ and $-6$. Dashed curves:
initial profile; solid curves: solutions at $t=100$. In all cases
$A=0.05$ and $\omega=0.2$; the axis limits are the same in all
panels.}
\label{fig:artificial_material_response}
\end{minipage}
\par\vspace{6mm}

Taken together, the computations show that, for the amplitudes of the
irregularity and the distance considered here, the character of the
evolution of a given pulse (propagation as a solitary wave,
disintegration into several solitary waves, or dispersion into a wave
train) is decided by the material through the sign and the magnitude
of $\mathcal{N}_0$, whereas the irregularity of the free surfaces modifies the
amplitudes of the individual waves by a few per cent and their
velocities by a fraction of a per cent. All amplitudes and positions
quoted in this section refer to the normalised variables
\eqref{eq:numerical_notation}.

\section{Conclusions}
\label{sec:conclusions}

The propagation of long nonlinear extensional waves in a compressible
hyperelastic layer with irregular, traction-free surfaces has been
studied within the framework of finite elasticity. The free surfaces were
taken to be symmetric with respect to the middle plane, and their
irregularity was allowed to change the local thickness by a finite
amount on a length scale long compared with the wavelength. In a
distinguished limit in which weak nonlinearity, dispersion and the
slow variation of the thickness appear at the same order, the method
of multiple scales applied to the two-dimensional boundary-value
problem led to a Korteweg--de Vries equation with variable
coefficients for the leading-order longitudinal strain.

The structure of this equation is simple and has a direct mechanical
interpretation. The velocity of the long wave is that of extensional
waves in a plate and is not affected by the irregularity, whose
influence is deferred to the order at which the amplitude evolves.
There, the geometry enters in only two ways: through a linear term
that expresses the conservation of the energy flux through a
cross-section, and through the factor $P^2$, the square of the local
thickness, multiplying the dispersive term. The nonlinear coefficient
is independent of the geometry; it is the coefficient of the
quadratic term in the uniaxial response of a layer with free surfaces
and involves the third-order elastic constants only through two
combinations. The dispersive coefficient, in contrast, depends on
Poisson's ratio alone and coincides with the long-wave limit of the
fundamental symmetric branch of the Rayleigh--Lamb frequency equation;
it arises from the inertia of the transverse contraction associated
with the longitudinal strain. The linear and the nonlinear elastic
properties of the material thus enter the long-wave model in clearly
separated ways.

After the linear geometric term was removed and the dispersive
coefficient was normalised, the effect of the material and of the
geometry was found to be carried entirely by a single coefficient,
$\mathcal{N}_0P^{-5/2}$. Its sign is fixed by the material; for iron
and copper it is such that compressive solitary waves of strain exist.
For special choices of the surface irregularity, namely a uniform
layer and a thickness varying as the cube root of the distance, the
normalised equation reduces to forms known to be integrable, the
Korteweg--de Vries and the cylindrical Korteweg--de Vries equations;
since the latter profile is admissible in a bounded region only, bounded
irregularities, and periodic ones in particular, require a numerical
treatment. The computations carried out for periodically irregular
layers of iron and copper and for model materials showed that the
type of evolution of a given pulse, namely propagation as a solitary
wave, disintegration into a train of solitary waves, or dispersion
into an oscillatory wave train, is decided by the material, while the
irregularity of the free surfaces redistributes the amplitudes of the
individual waves and changes their velocities only slightly.

Several extensions of the present work suggest themselves. The
response of a solitary wave to a localized change of thickness, such
as a step or a bump, is of interest in its own right: an abrupt
thinning is expected to produce secondary solitary waves, and an
abrupt thickening a dispersive tail together with a shelf behind the
transmitted wave, and these features can be described by combining the
inverse scattering theory of the uniform layer with the adiabatic
theory of slowly varying solitary waves
\cite{GrimshawPudjaprasetya2004,GrimshawYuan2016}. The formation and the
dynamics of the trailing shelf generated by a slowly varying thickness
deserve a separate analysis \cite{ElGrimshaw2002,Zhi2021}. For step-like initial data, and for
pulses of the polarity opposite to that of the solitary waves, the
evolution leads to dispersive shock waves, whose structure in a layer
of variable thickness could be studied by means of the Whitham
modulation theory with variable coefficients
\cite{ElGrimshaw2002,Ablowitz2016}. Finally, the case in
which the quadratic nonlinearity vanishes, the antisymmetric (flexural)
mode, layers with non-symmetric free surfaces, and layers bonded to a
substrate \cite{DeliktasTeymur2022,Sezer2026} call for different scalings and lead to evolution equations
of other types.

\section*{CRediT authorship contribution statement}
\textbf{Ali Demirci:} Writing – review and editing, Writing-original draft, Visualization, Investigation, Software, Formal analysis, Funding acquisition. \textbf{Haldun Taha Uzunoglu:} Writing – review and editing, Writing-original draft, Visualization, Investigation, Software, Formal analysis, Methodology.
\section*{Decleration of competing interest}
The authors declare that they have no known competing financial interests or personal relationships that could have appeared to influence the work reported in this paper.
\section*{Acknowledgements}
A. D. was supported under Grant No. TGA-2025-46559  by Istanbul Technical University Office of Scientific Research Projects (ITU BAPSIS). This study constitutes part of the Ph.D. thesis of H. T. U.
at Istanbul Technical University.

\section*{Data availability}
No data was used for the research described in the article.

\bibliographystyle{model1-num-names}
\bibliography{cas-refs}

\appendix

\section{Stress components}
\label{app:constitutive}

For the plane motion \eqref{eq:motion}, the strain energy function of
Section~\ref{subsec:constitutive} and the relation \eqref{eq:hyperelastic}
give, up to and including terms of the second degree in the
displacement gradients,
\begin{align}
\frac{T_{11}}{\rho_0}
&=
c_L^2u_{1,X}
+\left(c_L^2-2c_T^2\right)u_{2,Y}
+\eta_1u_{1,X}^2
+\eta_2\left(u_{2,Y}^2+2u_{1,X}u_{2,Y}\right)
+\eta_3u_{1,Y}u_{2,X}
+\eta_4\mathcal{K}_1,
\label{app:T11}
\\
\frac{T_{21}}{\rho_0}
&=
c_T^2\left(u_{1,Y}+u_{2,X}\right)
+\left(\eta_3u_{2,X}+2\eta_4u_{1,Y}\right)\mathcal{K}_2,
\label{app:T21}
\\
\frac{T_{12}}{\rho_0}
&=
c_T^2\left(u_{1,Y}+u_{2,X}\right)
+\left(2\eta_4u_{2,X}+\eta_3u_{1,Y}\right)\mathcal{K}_2,
\label{app:T12}
\\
\frac{T_{22}}{\rho_0}
&=
c_L^2u_{2,Y}
+\left(c_L^2-2c_T^2\right)u_{1,X}
+\eta_1u_{2,Y}^2
+\eta_2\left(u_{1,X}^2+2u_{1,X}u_{2,Y}\right)
+\eta_3u_{1,Y}u_{2,X}
+\eta_4\mathcal{K}_1,
\label{app:T22}
\end{align}
where
\begin{equation}
\mathcal{K}_1=u_{1,Y}^2+u_{2,X}^2,
\qquad
\mathcal{K}_2=u_{1,X}+u_{2,Y},
\label{app:K1K2}
\end{equation}
and $c_L$, $c_T$ and $\eta_i$ are defined by
\eqref{eq:wave_speeds}--\eqref{eq:beta_definition}. In components,
the balance equations \eqref{eq:balance_momentum} are
\begin{equation}
T_{11,X}+T_{21,Y}=\rho_0\ddot{u}_1,
\qquad
T_{12,X}+T_{22,Y}=\rho_0\ddot{u}_2,
\label{app:balance_components}
\end{equation}
and the traction conditions \eqref{eq:traction_free} on the upper
free surface $Y=H(X)$ are
\begin{equation}
-H'(X)\,T_{11}+T_{21}=0,
\qquad
-H'(X)\,T_{12}+T_{22}=0 .
\label{app:traction_components}
\end{equation}

\section{Dimensionless operators}
\label{app:operators}

Equations \eqref{app:T11}--\eqref{app:traction_components} are
expressed in the variables \eqref{eq:dimensionless_variables} by means
of \eqref{eq:scaled_gradients} and
\eqref{eq:epsilon_delta_relation}. The first balance equation and the
first traction condition are multiplied by $\varepsilon L/\rho_0$ and
$\varepsilon/(\delta\rho_0)$, and the second ones by
$\varepsilon^2L/(\delta\rho_0)$ and $\varepsilon/\rho_0$,
respectively, so that the powers of $\varepsilon$ in the operators
below coincide with the orders of the perturbation problems of
Section~\ref{sec:multiple_scales}. With
\begin{equation}
k_1=\left(\frac{\partial u}{\partial y}\right)^2
+\varepsilon^2\left(\frac{\partial v}{\partial x}\right)^2,
\qquad
k_2=\frac{\partial u}{\partial x}+\frac{\partial v}{\partial y},
\label{app:k1k2}
\end{equation}
which are the dimensionless forms of $\varepsilon\mathcal{K}_1$ and
$\mathcal{K}_2$, and with subscripts $x$, $y$ and $\tau$ denoting
partial derivatives, the linear operators are
\begin{align}
\mathcal{L}_1(u,v)
&=
\varepsilon c^2u_{\tau\tau}
-\varepsilon c_L^2u_{xx}
-c_T^2u_{yy}
-\varepsilon\left(c_L^2-c_T^2\right)v_{xy},
\label{app:L1}
\\
\mathcal{L}_2(u,v)
&=
\varepsilon^2c^2v_{\tau\tau}
-\varepsilon^2c_T^2v_{xx}
-\varepsilon c_L^2v_{yy}
-\varepsilon\left(c_L^2-c_T^2\right)u_{xy},
\label{app:L2}
\end{align}
and the nonlinear operators are
\begin{align}
\mathcal{N}_1(u,v)
={}&
\frac{\partial}{\partial y}
\left[\left(2\eta_4u_y+\varepsilon\eta_3v_x\right)k_2\right]
+\varepsilon\frac{\partial}{\partial x}
\left[
\eta_1u_x^2+\eta_2v_y^2+2\eta_2u_xv_y+\eta_3u_yv_x
\right]
+\eta_4\frac{\partial k_1}{\partial x},
\label{app:N1}
\\
\mathcal{N}_2(u,v)
={}&
\varepsilon\frac{\partial}{\partial x}
\left[\left(2\varepsilon\eta_4v_x+\eta_3u_y\right)k_2\right]
+\varepsilon\frac{\partial}{\partial y}
\left[
\eta_2u_x^2+\eta_1v_y^2+2\eta_2u_xv_y+\eta_3u_yv_x
\right]
+\eta_4\frac{\partial k_1}{\partial y}.
\label{app:N2}
\end{align}
On the upper free surface $y=P$, where
$\partial P/\partial x=\varepsilon P'(x_1)$, the traction operators are
\begin{align}
\mathcal{B}^{+}_1(u,v)
={}&
c_T^2\left(u_y+\varepsilon v_x\right)
+\left(2\eta_4u_y+\varepsilon\eta_3v_x\right)k_2
\nonumber\\
&-\varepsilon P'
\Big[
\varepsilon\left\{
c_L^2u_x+\left(c_L^2-2c_T^2\right)v_y
+\eta_1u_x^2+\eta_2v_y^2+2\eta_2u_xv_y+\eta_3u_yv_x
\right\}
+\eta_4k_1
\Big],
\label{app:B1plus}
\\
\mathcal{B}^{+}_2(u,v)
={}&
\varepsilon\left\{
c_L^2v_y+\left(c_L^2-2c_T^2\right)u_x
+\eta_1v_y^2+\eta_2u_x^2+2\eta_2u_xv_y+\eta_3u_yv_x
\right\}
+\eta_4k_1
\nonumber\\
&-\varepsilon^2P'
\Big[
c_T^2\left(u_y+\varepsilon v_x\right)
+\left(\eta_3u_y+2\varepsilon\eta_4v_x\right)k_2
\Big].
\label{app:B2plus}
\end{align}
The operators $\mathcal{B}^{-}_i$ on the lower free surface $y=-P$ are obtained
from \eqref{app:B1plus} and \eqref{app:B2plus} by replacing $P'$ with
$-P'$. In all these expressions $c_L$, $c_T$, $c$ and $\eta_i$ retain
their dimensions, every term having the dimension of a squared
velocity.

\section{Operators of the perturbation problems}
\label{app:order_operators}

With \eqref{eq:ms_derivative_expansion} and \eqref{eq:ms_expansion},
the coefficients of $\varepsilon^2$ and $\varepsilon^3$ in
\eqref{app:L1}--\eqref{app:B2plus} define the following operators. In
the second-order problem,
\begin{align}
\mathcal{L}^{21}
&=
c^2\frac{\partial^2u_1}{\partial\tau^2}
-c_L^2\frac{\partial^2u_1}{\partial x_0^2}
-\left(c_L^2-c_T^2\right)
\frac{\partial^2v_1}{\partial x_0\partial y},
\\
\mathcal{L}^{22}
&=
-c_L^2\frac{\partial^2v_1}{\partial y^2}
-\left(c_L^2-c_T^2\right)
\frac{\partial^2u_1}{\partial x_0\partial y},
\\
\mathcal{N}^{21}
&=
\eta_4\frac{\partial}{\partial x_0}
\left[\left(\frac{\partial u_1}{\partial y}\right)^2\right]
+\frac{\partial}{\partial y}
\left[
2\eta_4\frac{\partial u_1}{\partial y}
\left(
\frac{\partial u_1}{\partial x_0}+\frac{\partial v_1}{\partial y}
\right)
\right],
\\
\mathcal{N}^{22}
&=
\eta_4\frac{\partial}{\partial y}
\left[\left(\frac{\partial u_1}{\partial y}\right)^2\right],
\\
\mathcal{B}^{21}
&=
c_T^2\frac{\partial v_1}{\partial x_0}
+2\eta_4\frac{\partial u_1}{\partial y}
\left(
\frac{\partial u_1}{\partial x_0}+\frac{\partial v_1}{\partial y}
\right),
\\
\mathcal{B}^{22}
&=
c_L^2\frac{\partial v_1}{\partial y}
+\left(c_L^2-2c_T^2\right)\frac{\partial u_1}{\partial x_0}
+\eta_4\left(\frac{\partial u_1}{\partial y}\right)^2 .
\end{align}
In the third-order problem,
\begin{align}
\mathcal{L}^{31}
={}&
c^2\frac{\partial^2u_2}{\partial\tau^2}
-c_L^2
\left(
\frac{\partial^2u_2}{\partial x_0^2}
+2\frac{\partial^2u_1}{\partial x_0\partial x_1}
\right)
-\left(c_L^2-c_T^2\right)
\left(
\frac{\partial^2v_1}{\partial x_1\partial y}
+\frac{\partial^2v_2}{\partial x_0\partial y}
\right),
\\
\mathcal{L}^{32}
={}&
c^2\frac{\partial^2v_1}{\partial\tau^2}
-c_T^2\frac{\partial^2v_1}{\partial x_0^2}
-c_L^2\frac{\partial^2v_2}{\partial y^2}
-\left(c_L^2-c_T^2\right)
\left(
\frac{\partial^2u_1}{\partial x_1\partial y}
+\frac{\partial^2u_2}{\partial x_0\partial y}
\right),
\\
\mathcal{N}^{31}
={}&
\frac{\partial}{\partial x_0}
\Bigg[
\eta_1\left(\frac{\partial u_1}{\partial x_0}\right)^2
+\eta_2\left(\frac{\partial v_1}{\partial y}\right)^2
+2\eta_2\frac{\partial u_1}{\partial x_0}\frac{\partial v_1}{\partial y}
+\eta_3\frac{\partial u_1}{\partial y}\frac{\partial v_1}{\partial x_0}
+2\eta_4\frac{\partial u_1}{\partial y}\frac{\partial u_2}{\partial y}
\Bigg]
+\eta_4\frac{\partial}{\partial x_1}
\left[\left(\frac{\partial u_1}{\partial y}\right)^2\right]
\nonumber\\
&+
\frac{\partial}{\partial y}
\Bigg[
2\eta_4\frac{\partial u_1}{\partial y}
\left(
\frac{\partial u_1}{\partial x_1}
+\frac{\partial u_2}{\partial x_0}
+\frac{\partial v_2}{\partial y}
\right)
+\left(
\eta_3\frac{\partial v_1}{\partial x_0}
+2\eta_4\frac{\partial u_2}{\partial y}
\right)
\left(
\frac{\partial u_1}{\partial x_0}+\frac{\partial v_1}{\partial y}
\right)
\Bigg],
\\
\mathcal{N}^{32}
={}&
\frac{\partial}{\partial y}
\Bigg[
\eta_2\left(\frac{\partial u_1}{\partial x_0}\right)^2
+\eta_1\left(\frac{\partial v_1}{\partial y}\right)^2
+2\eta_2\frac{\partial u_1}{\partial x_0}\frac{\partial v_1}{\partial y}
+\eta_3\frac{\partial u_1}{\partial y}\frac{\partial v_1}{\partial x_0}
+2\eta_4\frac{\partial u_1}{\partial y}\frac{\partial u_2}{\partial y}
\Bigg]
\nonumber\\
&+
\frac{\partial}{\partial x_0}
\Bigg[
\eta_3\frac{\partial u_1}{\partial y}
\left(
\frac{\partial u_1}{\partial x_0}+\frac{\partial v_1}{\partial y}
\right)
\Bigg],
\\
\mathcal{B}^{31}
={}&
c_T^2
\left(
\frac{\partial v_2}{\partial x_0}+\frac{\partial v_1}{\partial x_1}
\right)
+2\eta_4\frac{\partial u_1}{\partial y}
\left(
\frac{\partial u_1}{\partial x_1}
+\frac{\partial u_2}{\partial x_0}
+\frac{\partial v_2}{\partial y}
\right)
+\left(
\eta_3\frac{\partial v_1}{\partial x_0}
+2\eta_4\frac{\partial u_2}{\partial y}
\right)
\left(
\frac{\partial u_1}{\partial x_0}+\frac{\partial v_1}{\partial y}
\right)
\nonumber\\
&-
P'
\Bigg[
c_L^2\frac{\partial u_1}{\partial x_0}
+\left(c_L^2-2c_T^2\right)\frac{\partial v_1}{\partial y}
+\eta_4\left(\frac{\partial u_1}{\partial y}\right)^2
\Bigg],
\\
\mathcal{B}^{32}
={}&
c_L^2\frac{\partial v_2}{\partial y}
+\left(c_L^2-2c_T^2\right)
\left(
\frac{\partial u_1}{\partial x_1}+\frac{\partial u_2}{\partial x_0}
\right)
+\eta_1\left(\frac{\partial v_1}{\partial y}\right)^2
+\eta_2\left(\frac{\partial u_1}{\partial x_0}\right)^2
+2\eta_2\frac{\partial u_1}{\partial x_0}\frac{\partial v_1}{\partial y}
\nonumber\\
&+
\eta_3\frac{\partial u_1}{\partial y}\frac{\partial v_1}{\partial x_0}
+2\eta_4\frac{\partial u_1}{\partial y}\frac{\partial u_2}{\partial y}
-c_T^2P'\frac{\partial u_1}{\partial y}.
\end{align}
All terms containing $\partial u_1/\partial y$ vanish identically once
the first-order solution \eqref{eq:ms_first_order_u_final} is used; they
are retained here so that the operators are valid for arbitrary
arguments.

\end{document}